\documentclass[aps,prc,superscriptaddress,twoside,twocolumn,nofootinbib,showpacs]{revtex4-1}
\usepackage{amsmath,amssymb}
\usepackage{mathtools}
\usepackage{bbold}
\usepackage[usenames]{color}
\usepackage{braket}
\usepackage{graphicx,bm}
\usepackage{slashed}
\usepackage{ulem}

\newcommand*{\DRarrow}{\mathrel{\rotatebox[origin=c]{180}{$\Lsh$}}}
\newcommand*{\abss}[1]{\left|#1\right|^2}
\newcommand*{\nunit}[1][n]{\hat{\bm{#1}}}

\allowdisplaybreaks

\begin{document}

\title{\boldmath Model-independent aspects of the reaction
$\bar{K} + N \to K + \Xi$}

\author{Benjamin C. Jackson}%
\affiliation{Department of Physics and Astronomy, The University of Georgia,
Athens, GA 30602, USA}

\author{Yongseok Oh}%
\email{yohphy@knu.ac.kr}
\affiliation{Department of Physics, Kyungpook National University,
Daegu 702-701, Korea}
\affiliation{Asia Pacific Center for Theoretical Physics, Pohang,
Gyeongbuk 790-784, Korea}

\author{H. Haberzettl}%
\email{helmut@gwu.edu}
\affiliation{Institute for Nuclear Studies and Department of Physics,
The George Washington University, Washington, DC 20052, USA}

\author{K. Nakayama}%
\email{nakayama@uga.edu}
\affiliation{Department of Physics and Astronomy, The University of Georgia,
Athens, GA 30602, USA}
\affiliation{Institut f\"ur Kernphysik and Center for Hadron Physics,
Forschungszentrum J\"ulich, 52425 J\"ulich, Germany}

\date{\today}

\begin{abstract}
Various model-independent aspects of the $\bar{K} N \to K \Xi$ reaction are
investigated, starting from the determination of the most general structure of
the reaction amplitude for $\Xi$ baryons with $J^P=\frac12^\pm$ and
$\frac32^\pm$ and the observables that allow a complete determination of these
amplitudes. Polarization observables are constructed in terms of spin-density
matrix elements. Reflection symmetry about the reaction plane is exploited, in
particular, to determine the parity of the produced $\Xi$ in a
model-independent way. In addition, extending the work of Biagi \textit{et
al.\/} [Z. Phys.\ C \textbf{34}, 175 (1987)], a way is presented of determining
simultaneously the spin and parity of the ground state of $\Xi$ baryon as well
as those of the excited $\Xi$ states.
\end{abstract}


\maketitle


\section{Introduction}

Multi-strangeness baryons have played an important role in the development of
our understanding of strong interactions. For example, the prediction and
discovery of the $\Omega$ baryon, with strangeness $S=-3$, has been a
spectacular confirmation of how well the SU(3) flavor symmetry works in strong
interactions. Nevertheless, more than a half century later, our knowledge of
multi-strangeness baryons is still very limited as can be seen from the fact
that the spin of the $\Omega^-(1672)$ was only recently
confirmed~\cite{BABAR06,PDG12}.

In this work, we concentrate on the $\Xi$ baryons with strangeness $S=-2$. So
far, studies of $\Xi$ physics have been very scarce.  The present situation can
be summarized as follows. (i) The SU(3) flavor symmetry allows as many $\Xi$
states as there are $N^*$ and $\Delta^*$ resonances combined ($\sim 44$).
However, until now, only 11 $\Xi$ resonances have been discovered~\cite{PDG12}.
(ii) Being  $S=-2$ baryons, if there are no strange particles in the initial
state, $\Xi$ are produced only indirectly and have relatively low production
rates. In fact, the yield is only of the order of nb in the photoproduction
reaction~\cite{CLAS07b}, whereas the yield is of the order of
$\mu$b~\cite{FMMR83} in the hadronic, $\bar{K}$-induced reaction (where the
$\Xi$ is produced directly because of the presence of an $S=-1$ $\bar{K}$ meson
in the initial state).

With the advent of new particle accelerators capable of reaching higher
energies and advances in technologies and experimental techniques, we are now
in a better position than ever to study multi-strangeness baryons. Indeed, the
CLAS Collaboration at the Thomas Jefferson National Accelerator Facility (JLab)
plans to initiate a $\Xi$ spectroscopy program using the upgraded 12-GeV
machine. The Collaboration is also expected to measure exclusive $\Omega$
photoproduction for the first time~\cite{VSC12}. Some data for the $\Xi$ ground
state are already available~\cite{CLAS07b} obtained from the 6-GeV
machine. In addition, J-PARC proposes to study the $\Xi$
baryons via the $\bar{K} N \to K \Xi$ and $\pi N \to K K \Xi$ reactions as well
as $\Omega$ production~\cite{Ahn06,Takahashi13}, and at the Facility for
Antiproton and Ion Research (FAIR) of GSI, the reaction $\bar{p} p \to
\bar{\Xi} \Xi$ will be studied~\cite{PANDA09}. For a more complete compilation
of baryon spectra, the $\Xi$ baryons should be studied as an integral part of
any baryon spectroscopy program.

Theoretical studies of the $\Xi$ baryons are hampered mainly by the scarcity of
experimental data. The existing theoretical models cannot be well constrained
and, as a consequence, there is strong model-dependence in predictions of the
$\Xi$ spectrum. In particular, one of the current open issues in the $\Xi$
spectrum concerns the low mass of the $\Xi(1690)$ and $\Xi(1620)$, i.e., the
nature of the third lowest $\Xi$ state~\cite{Oh07}. Here, different approaches,
such as the non-relativistic and relativistic quark
models~\cite{CIK81,CI86,PR07}, one-boson-exchange model~\cite{GR96b}, large
$N_c$ model~\cite{CC00,SGS02,GSS03,MS04b,MS06b}, QCD sum
rules~\cite{LL02,JO96}, and Skyrme model~\cite{Oh07}, yield contradictory
predictions for the nature of these resonances. The planned new experimental
studies as mentioned above are expected to play a key role in addressing such
open problems. Quite recently, lattice QCD calculations of the baryon spectra,
including those of $\Xi$ and $\Omega$ baryons, have been
reported~\cite{HSC12,BGR13}.

To extract relevant information on $\Xi$ resonances from the experimental data,
a reliable reaction model is required. To date, for photoproduction reactions,
there exist currently only the work of some of the present
authors~\cite{NOH06,MON11} analyzing the available CLAS data~\cite{CLAS07b}. In
$\bar{K}$-induced reactions, recent calculations are reported by Sharov
\textit{et al.\/}~\cite{SKL11} and by Shyam \textit{et al.\/}~\cite{SST11}.
Thus, further theoretical studies on this subject are timely for suggesting
directions to experimental studies by providing predictions on the $\Xi$ baryon
production processes and for giving tools to analyze the forthcoming data. One
feature of these production processes is that the $t$-channel processes of an
intermediate meson production are suppressed since the exchanged intermediate
meson should be exotic having two units of strangeness. As a consequence, the
production of a $\Xi$ state is dominated by  intermediate $S=-1$ hyperons.
Therefore, by analyzing the production mechanisms of the $\Xi$, one also hopes
to gain some insight into the spectrum and couplings of the $S = -1$ hyperons.

Our ongoing efforts to understand better the production process of $\Xi$
baryons are pursued along two lines of inquiry. On the one hand, to build
understanding of the dynamics of reactions involving strange particles, we are
engaged in model-dependent analyses within an effective Lagrangian approach
along the lines employed in the photoprocesses reported by some of us in
Refs.~\cite{NOH06,MON11}. Further results of such model-dependent analyses will
be reported elsewhere. In the present work, on the other hand, we report on
studying \textit{model-independent} aspects  of production processes of  $\Xi$
baryons  exploiting, in particular, some basic symmetries of the reactions in
question to determine the spin and parity quantum numbers of the $\Xi$
resonances.

The present paper is organized as follows. In Sec.~\ref{sec:Kbar}, following
the method of Ref.~\cite{NL05},  the most general structure of the reaction
amplitude for $\bar{K} N \to K \Xi$ is derived for a $\Xi$ of spin-1/2 or of
spin-3/2, and a set of observables is identified that determine the reaction
amplitude completely. Here, the reflection symmetry about the reaction plane is
exploited to determine the parity of the $\Xi$ resonances in a
model-independent manner. Furthermore, the coefficients that multiply each spin
structure in the reaction amplitude are expressed in terms of partial-wave
matrix elements, which readily allows for a partial-wave analysis when
sufficient data become available. Since the determination of the basic quantum
numbers of the produced $\Xi$ involve its spin observables, the
spin-density-matrix (SDM) approach is used, in Sec.~\ref{sec:SDMA}, to discuss
all relevant observables in terms of SDM elements. The SDM formalism is also
very convenient when dealing with high-spin $\Xi$ resonance productions.
In fact, the SDM approach has been often used in the description of
higher-spin particle production processes such as the (spin-1) vector meson
productions \cite{KCT98,KT00}.
Then, following Refs.~\cite{Chung71,BBBB87b}, the SDM elements are extracted
from the moments associated with the decay processes of the produced $\Xi$ in
conjunction with the self-analyzing properties of the ground state $\Xi$ and
$\Lambda$ in weak decays.
These moments, in turn, can be extracted from the measured angular
distribution of the decay product. Section~\ref{sec:summary} contains a summary
and some technical details of the derivations are given in several Appendices.

\section{\boldmath Structure of the $\bar{K} N \to K \Xi$ reaction amplitude}
\label{sec:Kbar}

In this section, we derive the most general structure of the amplitude for the
reaction of
\begin{equation}
\bar{K}(q)  + N(p) \to K(q')  + \Xi(p')  ~,
\label{eq:Kbar-1}
\end{equation}
following the method used in Ref.~\cite{NL05}. In the present work, we consider
the production of $\Xi$ of spin-1/2 and \mbox{-3/2}  with both positive and
negative parities. The method is quite general and, in principle, can be
applied to extract the spin structure of any reaction amplitude. In the above
equation, the arguments denote the four-momenta of the respective particles.

\begin{figure*}[t!]\centering
\includegraphics[width=0.9\textwidth,clip=]{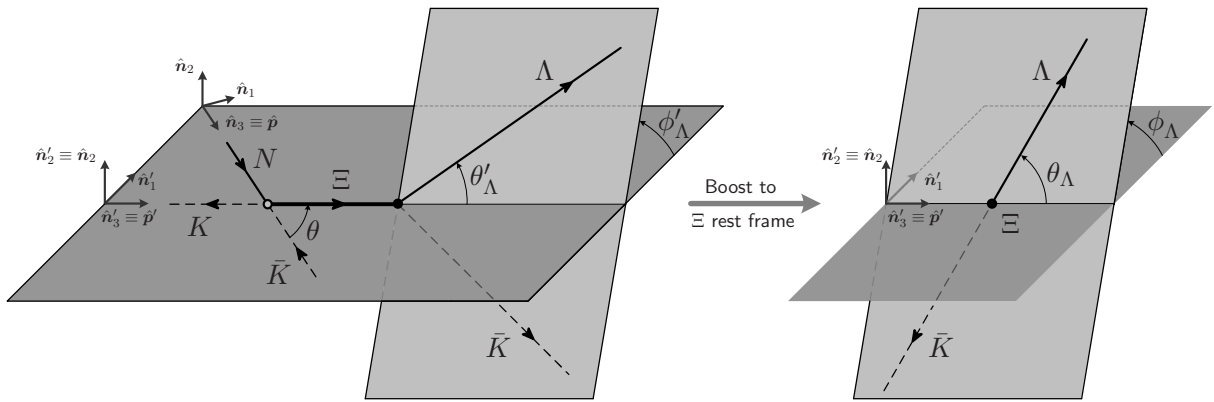}
\caption{\label{fig:1}%
 Coordinate systems used in describing the $\Xi$ production reaction and its
subsequent decay process. On the left, the production reaction  $\bar{K}N\to
K\Xi$ is shown in its center-of-momentum (CM) frame. The corresponding reaction
plane (indicated in dark gray) contains the nucleon and $\Xi$ momenta
${\bm{p}}$ and ${\bm{p}'}$, respectively. The basis vectors $\{ \nunit^{}_1,
\nunit^{}_2, \nunit^{}_3 \}$ are defined in Eq.~(\ref{eq:coord-syst}), with
$\nunit_3$ aligned with the nucleon momentum $\bm{p}$ and $\nunit_2$
perpendicular to the reaction plane; $\theta$ indicates the $\Xi$ emission
angle. The (primed) frame $\{\nunit'_1,\nunit'_2,\nunit'_3\}$ is obtained from
$\{\nunit^{}_1,\nunit^{}_2,\nunit^{}_3\}$ by rotating the latter about the
$\nunit^{}_2$ axis by $\theta$, which aligns $\nunit'_3$ with $\bm{p}'$ and
leaves $\nunit'_2 \equiv \nunit^{}_2$. The (light gray) plane tilted by the
angle $\phi'^{}_\Lambda$ about the $\nunit'_3\equiv\nunit[p]'$ axis is spanned
by the momenta of the decay products $\Lambda$ and $\bar{K}$. The polar and
azimuthal angles of the decay product $\Lambda$ in the rotated (primed) CM
frame are indicated by $\theta'_\Lambda$ and $ \phi'_\Lambda$, respectively. In
the boosted frame on the right, the decay process of the produced $\Xi$ at rest
is described in the $\{ \nunit'_1, \nunit'_2, \nunit'_3 \}$ coordinate system.
The polar and azimuthal angles of the decay product $\Lambda$ are indicated
here by $\theta_\Lambda$ and $\phi^{}_\Lambda$, respectively. For the latter
angle, one has $\phi^{}_\Lambda\equiv \phi'^{}_\Lambda$ since the boost happens
along the corresponding tilt axis.}
\end{figure*}

The reaction in Eq.~(\ref{eq:Kbar-1}) is described in its center-of-momentum
(CM) frame, where $\bm{q} = - \bm{p}$ and $\bm{q}' = - \bm{p}'$. For further
convenience, we define the three mutually orthogonal unit vectors $\nunit_i^{}$
$(i=1,2,3)$ in terms of the independent momenta available in the reaction, i.e.,
\begin{subequations}\label{eq:coord-syst}
\begin{align}
\nunit_1^{}  & \equiv \frac{(\bm{p} \times \bm{p}') \times \bm{p}}
{| (\bm{p} \times \bm{p}') \times \bm{p} |}~  ,
\\
\nunit_2^{} & \equiv \frac{ \bm{p} \times \bm{p}' }
{| \bm{p} \times \bm{p}'  |} ~ ,
\\
\nunit_3^{} & \equiv \frac{\bm{p}}{| \bm{p} |}  ~,
\end{align}
\end{subequations}
where $\bm{p}$ and $\bm{p}'$ denote the three-momenta of the nucleon and $\Xi$,
respectively. Note that $\bm{p}$ and $\bm{p}'$ define the reaction plane, such
that $\nunit_2^{}$ is perpendicular to the reaction plane. The coordinate-system
setup is shown in Fig.~\ref{fig:1}. Throughout this paper, the hat notation for
vectors is used to indicate unit vectors, i.e., $\nunit[a] \equiv
\bm{a}/|\bm{a}|$ for an arbitrary vector $\bm{a}$.
The quantization axis is chosen to be along $\nunit_3^{}$.
We also use the alternative Cartesian notation $i=x,y,z$ for the indices of
the unit vectors $\nunit_i^{}$.

\subsection{\boldmath Production of $\Xi$ with $J^P=\frac12^\pm$}
\label{subsec:Kbar-12}

First, we consider spin-parity $J^P=\frac12^\pm$ for the $\Xi$ produced in
reaction~(\ref{eq:Kbar-1}). Following the method of Ref.~\cite{NL05}, the most
general spin structure of the reaction amplitude, consistent with basic
symmetries, is
\begin{subequations}\label{eq:Kbar-12-a1}
\begin{align}
\hat{M} &= M'_0+ M'_2 \, \bm{\sigma} \cdot (\hat{\bm{p}} \times \hat{\bm{p}}')~,
                                            \quad & \mbox{for } J^P=\tfrac 12^+ ~,
\label{eq:Kbar-12-a1+}\\
\hat{M} &=M'_1 \, \bm{\sigma} \cdot \hat{\bm{p}}'
+ M'_3 \, \bm{\sigma} \cdot \hat{\bm{p}}~, \quad & \mbox{for }  J^P=\tfrac 12^- ~,
\label{eq:Kbar-12-a1-}
\end{align}
\end{subequations}
where $\bm{\sigma} = (\sigma_1^{}, \sigma_2^{}, \sigma_3^{})$ stands for the vector
built up of usual Pauli spin operators.%
\footnote{Note that the spin structure for the positive-parity $\Xi$ in
Eq.~(\ref{eq:Kbar-12-a1+}) is identical to the familiar structure of the $\pi N$
elastic scattering amplitude. However, obviously, the isospin structure is
different.} Equations~(\ref{eq:Kbar-12-a1+}) and (\ref{eq:Kbar-12-a1-}) are
direct consequences of the amplitude's reflection symmetry about the reaction
plane~\cite{JW59,Bohr59}, which is further exploited in our analysis presented
below. Note that the coefficients $M'_1$ and $M'_2$ do not contain $S$-wave in
the final state.

For further convenience, we rewrite Eq.~(\ref{eq:Kbar-12-a1}) as
\begin{subequations}\label{eq:Kbar-12-a}
\begin{align}
\hat{M} &= M_0+ M_2 \, \bm{\sigma} \cdot \nunit_2^{}~,
\quad & \textstyle \mbox{for }  J^P=\frac 12^+ ~,  \\
\hat{M} &=M_1\, \bm{\sigma} \cdot \nunit_1^{} +
M_3\, \bm{\sigma} \cdot \nunit_3^{}~, \quad &\textstyle
\mbox{for } J^P=\frac 12^- ~,
\end{align}
\end{subequations}
using $\nunit[p]' = \cos\theta\, \nunit_3^{} + \sin\theta\, \nunit_1^{}$ and
$\nunit_3^{} = \nunit[p]$. The respective coefficients in
Eqs.~(\ref{eq:Kbar-12-a1}) and (\ref{eq:Kbar-12-a}) are related by
\begin{subequations}\label{eq:Kbar-12-a2}
\begin{align}
M'_0 & = M_0 ~,  &
M'_2 &= \frac{1}{\sin\theta} \, M_2 ~,  \\
M'_1 & =  \displaystyle\frac{1}{\sin\theta}  \, M_1  ~, &
M'_3 &= M_3 - \frac{\cos\theta}{\sin\theta} \, M_1  ~.
\end{align}
\end{subequations}
Following Ref.~\cite{NL05}, one may also express these coefficients in terms of
partial-wave matrix elements. The corresponding results are given in
Appendix~\ref{app:A}, which show, in particular, that the coefficients $M_1$
and $M_2$ vanish identically for $\Xi$ scattering angles $\theta = 0$ and
$\pi$, as can be seen in Eq.~(\ref{eq:A2}). The partial-wave expansions will
become particularly relevant once sufficient experimental data become available
to permit their full-fledged partial-wave analysis.
The isospin structure of the amplitudes in
Eq.~(\ref{eq:Kbar-12-a}) [or in Eq.~(\ref{eq:Kbar-12-a1})] is contained in the
coefficients $M_i$ as given explicitly by Eq.~(\ref{eq:A2}) in
Appendix~\ref{app:A}.

Once the spin structure of the reaction amplitude is determined, all
the observables can be readily expressed in terms of the amplitudes $M_i$
multiplying each spin structure. For the reaction under consideration, apart
from the cross section ($d\sigma/d\Omega$), a complete set of observables
includes the target asymmetry ($T$), recoil $\Xi$ polarization ($P$), and the
spin-transfer coefficient ($K$). For arbitrary spin orientations along
directions $\nunit[a]$ and $\nunit[b]$, their coordinate-independent
expressions are
\begin{subequations}\label{eq:Kbar-12-b-alt}
\begin{align}
\frac{d\sigma}{d\Omega} & \equiv \frac12 \,\mbox{Tr} [\hat{M} \hat{M}^\dagger] ~,
 \\
\frac{d\sigma}{d\Omega} T_a & \equiv \frac12
\,\mbox{Tr} [\hat{M}  \,\bm{\sigma}\cdot\nunit[a]\,\hat{M}^\dagger]  ~,  \\
\frac{d\sigma}{d\Omega} P_a & \equiv
\frac12 \,\mbox{Tr} [\hat{M} \hat{M}^\dagger  \,\bm{\sigma}\cdot\nunit[a]]  ~,  \\
\frac{d\sigma}{d\Omega} K_{ba} & \equiv
\frac12 \,\mbox{Tr} [\hat{M}  \,\bm{\sigma}\cdot\nunit[b]\, \hat{M}^\dagger  \,
\bm{\sigma}\cdot\nunit[a]]  ~.
\label{eq:Kbar-12-b-alt-d}
\end{align}
\end{subequations}%
For Cartesian directions  $\nunit^{}_i$ enumerated by $i =1,2,3$ $(=x,y,z)$, in
particular, one obtains
\begin{subequations}\label{eq:Kbar-12-b}
\begin{align}
\frac{d\sigma}{d\Omega} T_i &= \frac12
\,\mbox{Tr} [\hat{M} \sigma_i^{} \hat{M}^\dagger]  ~,  \\
\frac{d\sigma}{d\Omega} P_i &=
\frac12 \,\mbox{Tr} [\hat{M} \hat{M}^\dagger \sigma_i^{}]  ~,  \\
\frac{d\sigma}{d\Omega} K_{ij} &=
\frac12 \,\mbox{Tr} [\hat{M} \sigma_i^{} \hat{M}^\dagger\sigma_j^{}]  ~.
\end{align}
\end{subequations}
Of course, the $T$, $P$, and $K$ observables for arbitrary directions in
Eq.~(\ref{eq:Kbar-12-b-alt}) can be expressed as linear combinations of the
specific Cartesian expressions given in Eq.~(\ref{eq:Kbar-12-b}).

Due to symmetries of the reaction, eight observables vanish identically, i.e.,
$T_i=P_i=K_{iy}=K_{yi}=0$ for $i=x,z$, and of the remaining eight, only four
are independent for a given parity, which completely determine the amplitudes
$M_i$ in Eq.~(\ref{eq:Kbar-12-a}). Indeed, for a positive-parity $\Xi$, we have
\begin{subequations}\label{eq:Kbar-12-c}
\begin{align}
\frac{d\sigma}{d\Omega} = \frac{d\sigma}{d\Omega}K_{yy} & = |M_0|^2 + |M_2|^2 ~,
\label{eq:Kbar-12-c-a}
\\
\frac{d\sigma}{d\Omega} K_{xx}  =  \frac{d\sigma}{d\Omega} K_{zz}
& =  |M_0|^2 - |M_2|^2   ~,     \\
\frac{d\sigma}{d\Omega} T_y  = \frac{d\sigma}{d\Omega} P_y
& = 2 \,\mbox{Re}\left[ M_2 M^{*}_0 \right]  ~,
 \\
\frac{d\sigma}{d\Omega} K_{xz}  =  - \frac{d\sigma}{d\Omega} K_{zx}
& = 2 \,\mbox{Im}\left[ M_2 M_0^* \right]  ~,
\end{align}
\end{subequations}
and for a negative-parity $\Xi$, we obtain
\begin{subequations}\label{eq:Kbar-12-d}
\begin{align}
\frac{d\sigma}{d\Omega} = -\frac{d\sigma}{d\Omega}K_{yy} & = |M_1|^2 + |M_3|^2 ~,
\label{eq:Kbar-12-d-a}
 \\
\frac{d\sigma}{d\Omega}K_{xx}  = - \frac{d\sigma}{d\Omega}K_{zz}
& =  |M_1|^2 - |M_3|^2   ~,
 \\
\frac{d\sigma}{d\Omega} T_y  = - \frac{d\sigma}{d\Omega} P_y
& = 2 \,\mbox{Im}\left[ M_3 M^{*}_1 \right]  ~,
 \\
\frac{d\sigma}{d\Omega} K_{xz} =  \frac{d\sigma}{d\Omega} K_{zx}
& = 2 \,\mbox{Re}\left[ M_3 M_1^* \right]  ~.
\end{align}
\end{subequations}%
The respective first two relations in the two equation sets determine the
magnitudes of the amplitudes $M_0$, $M_2$ and $M_1$, $M_3$, respectively,
whereas the respective last two relations determine their phase differences.
Therefore, apart from an irrelevant overall phase, the observables in
Eqs.~(\ref{eq:Kbar-12-c}) and (\ref{eq:Kbar-12-d}) determine the amplitudes
$M_i$, $i=0,\ldots,3$, unambiguously. These results reveal that it is
experimentally demanding to determine the reaction amplitude completely, for it
requires measuring both the single- and double-polarization observables.

Comparing Eqs.~(\ref{eq:Kbar-12-c-a}) and (\ref{eq:Kbar-12-d-a}), one obtains
\begin{equation}
K_{yy}  =  \pi_\Xi^{}  ~,
\label{eq:Kbar-12-e}
\end{equation}
where $\pi_\Xi^{}$ stands for the parity of the produced $\Xi$. This result
is actually a direct consequence of reflection symmetry, as exploited in Bohr's
theorem~\cite{Bohr59,Satchler} and applied in Ref.~\cite{NOH12}. It,
therefore, provides a model-independent way of determining the parity of the
$\Xi$ resonance. Alternative expressions extracted from
Eqs.~(\ref{eq:Kbar-12-c}) and (\ref{eq:Kbar-12-d}) are~\cite{NOH12}
\begin{equation}
T_y = \pi_\Xi^{} \, P_y~,
\label{eq:Kbar-12-f}
\end{equation}
which involves only single polarization observables and
\begin{equation}
K_{xx} = \pi_\Xi^{} \, K_{zz}
\quad\text{and}\quad
K_{xz} = - \pi_\Xi^{} \,K_{zx}~.
\end{equation}
These results are all consequences of the reflection symmetry about the
reaction plane.

In Sec.~\ref{sec:SDMA}, we will perform the analysis in terms of the SDM
elements, which are equivalent to the observables discussed here.
The SDM elements are convenient quantities when dealing with spin observables,
especially when higher-spin particles are produced in the reaction.
They can be extracted from the information on the subsequent decay processes
of the produced particles, in conjunction with the self-analyzing property
of the decaying particles via a weak decay,
without the explicit measurement of the spin
polarizations of these produced particles.

\subsection{\boldmath Production of $\Xi$ with $J^P=\frac32^\pm$}

We now turn to the spin-parity $J^P=\frac32^\pm$. Again, following
Ref.~\cite{NL05}, the most general spin structure of the reaction amplitude is
given by
\begin{subequations}\label{eq:Kbar-32-b}
\begin{align}
\hat M &= F'_1\; \bm{T}^\dagger \cdot (\hat{\bm{p}} \times \hat{\bm{p}}')
+ F'_2\; \bm{T}^\dagger \cdot \hat{\bm{p}}' \, \bm{\sigma} \cdot \hat{\bm{p}}'
\nonumber\\
&\quad\mbox{}
+F'_3\, \left[ \bm{T}^\dagger \cdot \hat{\bm{p}} \, \bm{\sigma} \cdot \hat{\bm{p}}'
+ \bm{T}^\dagger \cdot \hat{\bm{p}}' \, \bm{\sigma} \cdot \hat{\bm{p}} \right]
\nonumber\\
&\quad\mbox{}
+ F'_4\;  \bm{T}^\dagger \cdot \hat{\bm{p}} \, \bm{\sigma} \cdot \hat{\bm{p}}
\\
\intertext{for $J^P=\frac 32^+$ and}
\hat{M} &= G'_1\, \left[ \bm{T}^\dagger \cdot \hat{\bm{p}} \,
\bm{\sigma} \cdot (\hat{\bm{p}} \times \hat{\bm{p}}')
+ \bm{T}^\dagger \cdot (\hat{\bm{p}} \times \hat{\bm{p}}') \,
\bm{\sigma} \cdot \hat{\bm{p}} \right]
\nonumber\\
&\quad\mbox{}
+ G'_2\, \left[ \bm{T}^\dagger \cdot \hat{\bm{p}}' \,
\bm{\sigma} \cdot (\hat{\bm{p}} \times \hat{\bm{p}}')
+ \bm{T}^\dagger \cdot (\hat{\bm{p}} \times \hat{\bm{p}}') \,
\bm{\sigma} \cdot \hat{\bm{p}}' \right]
\nonumber\\
&\quad\mbox{}
+ G'_3\; \bm{T}^\dagger \cdot \hat{\bm{p}}'
+ G'_4\; \bm{T}^\dagger \cdot \hat{\bm{p}}
\end{align}
\end{subequations}
for $J^P=\frac 32^-$. Here, $\bm{T}^\dagger$ stands for the (spin-1/2 $\to$
spin-3/2) transition operator. Its explicit representation may be found
elsewhere~\cite{ONL04}. In contrast to the spin-1/2 case, each parity of the
spin-3/2 case has four independent amplitudes, $F'_i$ and $G'_i$ ($i=1, \dots,
4$), respectively, and one needs at least eight independent observables to
determine them completely apart from an irrelevant overall phase. From the
above equations, it is obvious that only $F'_4$ and $G'_4$ contain an $S$-wave
in the final state. Also, $F'_2$ and $G'_2$ contain only $D$- and higher-waves
in the final state.

The amplitudes in Eq.~(\ref{eq:Kbar-32-b}) can be also rewritten as
\begin{subequations}\label{eq:Kbar-32-a}
\begin{align}
\hat M &= F_1\; \bm{T}^\dagger \cdot \nunit^{}_2
+ F_2\; \bm{T}^\dagger \cdot \nunit^{}_1 \, \bm{\sigma} \cdot \nunit^{}_1
\nonumber\\
&\quad\mbox{}
+ F_3\, \left[ \bm{T}^\dagger \cdot \nunit^{}_3 \, \bm{\sigma} \cdot \nunit^{}_1
+  \bm{T}^\dagger \cdot \nunit^{}_1 \, \bm{\sigma} \cdot \nunit^{}_3 \right]
\nonumber\\
&\quad\mbox{}
+ F_4\;  \bm{T}^\dagger \cdot \nunit^{}_3 \, \bm{\sigma} \cdot \nunit^{}_3
\\
\intertext{for $J^P=\frac 32^+$ and}
\hat{M} &= G_1\, \left[
\bm{T}^\dagger \cdot \nunit^{}_3 \, \bm{\sigma} \cdot \nunit^{}_2
+ \bm{T}^\dagger \cdot \nunit^{}_2 \, \bm{\sigma} \cdot \nunit^{}_3 \right]
\nonumber\\
&\quad\mbox{}
+ G_2\, \left[ \bm{T}^\dagger \cdot \nunit^{}_1 \, \bm{\sigma} \cdot \nunit^{}_2
+ \bm{T}^\dagger \cdot \nunit^{}_2 \, \bm{\sigma} \cdot \nunit^{}_1 \right]
\nonumber\\
&\quad\mbox{}
+ G_3\; \bm{T}^\dagger \cdot \nunit^{}_1
+ G_4\; \bm{T}^\dagger \cdot \nunit^{}_3
\end{align}
\end{subequations}
for $J^P=\frac 32^-$. The coefficients $F_i$ and $G_i$ are expressed in terms
of the partial-wave matrix elements as given in Appendix~\ref{app:A}. They are
also related to the corresponding coefficients $F'_i$ and $G'_i$ in
Eq.~(\ref{eq:Kbar-32-b}) by
\begin{subequations}\label{eq:Kbar-32-b1}
\begin{align}
F'_1 &= \frac{1}{\sin\theta} F_1~,
 \\
F'_2 &=  \frac{1}{\sin^2\theta} F_2  ~,
 \\
F'_3  &= \frac{1}{\sin\theta} F_3 - \frac{\cos\theta}{\sin^2\theta} F_2  ~,
 \\
F'_4  &= F_4 + \frac{\cos^2\theta}{\sin^2\theta} F_2
- 2\frac{\cos\theta}{\sin\theta} F_3 ~,
 \\
\intertext{and}
G'_1 &= \frac{1}{\sin\theta}G_1 - \frac{\cos\theta}{\sin^2\theta} G_2  ~,
 \\
G'_2 &=  \frac{1}{\sin^2\theta} G_2 ~,
 \\
G'_3 &= \frac{1}{\sin\theta} G_3 ~,
 \\
G'_4 &= G_4 - \frac{\cos\theta}{\sin\theta} G_3  ~.
\end{align}
\end{subequations}
The polarization observables for this case will be discussed in the next
section in terms of the SDM elements.

\section{Spin Density Matrix Approach}
\label{sec:SDMA}

As mentioned before, when dealing with higher-spin $\Xi$ (i.e., spins higher
than 1/2) in particular, it is more convenient to continue the analysis of the
$\bar{K} N \to K \Xi$ reaction in terms of spin-density matrix (SDM) elements.
A similar (but not identical) analysis to the present one based on the SDM
formalism was performed in Ref.~\cite{Deen} for a general two-body reaction
with unpolarized initial state.
Also, the reaction $\bar{K} N \to \omega \Lambda$ was analyzed within the SDM
approach in Ref.~\cite{BBDL72}.

In Sec.~\ref{sec:Kbar}, we have exploited the mirror (or reflection) symmetry
about the reaction plane in our analysis, in particular, for the parity
determination of the $\Xi$ resonances. In fact, as long as the production
process conserves total parity, the reaction amplitude should have this
symmetry~\cite{JW59,Bohr59}. This mirror operation is equivalent to doing a
parity transformation followed by a subsequent rotation by $180^\circ$ about
the $\nunit_2^{}$-axis: $\hat{\cal{P}}_y=\hat R_y(180^\circ)\hat{\cal{P}}$. The
resulting symmetry, in terms of the spin matrix element, is
\begin{eqnarray}
\braket{S_f \, m_f^{} | \hat M | S_i\, m_i^{}} &=&
\braket{S_f \, m_f^{} | \hat{\mathcal{P}}_y^\dagger \mathcal{P}_y \hat M
\mathcal{P}^\dagger_y \mathcal{P}_y | S_i\, m_i^{}}
\nonumber \\
&=& \pi_f^{} \pi_i^{} \, (-1)^{(S_f - m_f^{}) - (S_i-m_i^{})}
\nonumber \\ && \mbox{} \times
\braket{S_f -m_f^{} | \hat M | S_i -m_i^{}}  ~,
\label{mirror_sym1}
\end{eqnarray}
and holds as long as the quantization axis is in the production plane. Here,
$\pi_{i(f)}$ is the intrinsic parity of the initial (final) state.

Based on this symmetry, the $J^P=\frac12^\pm$  $\Xi$ production amplitude,
$\hat M$ given by Eq.~(\ref{eq:Kbar-12-a}), is completely described by two
complex helicity amplitudes, $\mathcal{H}_1$ and $\mathcal{H}_2$, given by
the spin matrix elements,
\begin{subequations}\label{M_1hf_hel}
\begin{align}
\mathcal{H}_1 &\equiv \bra{\lambda_\Xi^{} = \tfrac{1}{2}} \hat{M}
\ket{\lambda_N^{} = \tfrac12}
\nonumber \\
&= \pi_\Xi^{} \bra{\lambda_\Xi^{} = -\tfrac12} \hat{M}
\ket{\lambda_N^{} = -\tfrac12} ~,
 \\[1ex]
\mathcal{H}_2 &\equiv \bra{\lambda_\Xi^{} = \tfrac12} \hat{M}
\ket{\lambda_N^{} = -\tfrac12}
\nonumber\\
&= -\pi_\Xi^{} \bra{\lambda_\Xi^{} = -\tfrac12} \hat{M}
\ket{ \lambda_N^{} = \tfrac12}  ~,
\end{align}
\end{subequations}
where $\lambda_N^{}$ and $\lambda_\Xi^{}$ denote the helicity of the initial
nucleon and final $\Xi$, respectively. Here, reference to the spin quantum
numbers $S_\Xi = S_N = 1/2$ has been suppressed. The helicity amplitudes are
related to the coefficient amplitudes in Eq.~(\ref{eq:Kbar-12-a}) by
\begin{subequations}\label{eq:HM-12+}
\begin{align}
\mathcal{H}_1 &=  M_0\cos\frac{\theta}{2}  + iM_2\sin\frac{\theta}{2}  ~,
 \\
\mathcal{H}_2 &= -i M_2 \cos\frac{\theta}{2} + M_0\sin\frac{\theta}{2}
\end{align}
\end{subequations}
for a positive parity $\Xi$, and by
\begin{subequations}\label{eq:HM-12-}
\begin{align}
\mathcal{H}_1 & = M_3\cos\frac{\theta}{2} + M_1\sin\frac{\theta}{2}  ~,
 \\
\mathcal{H}_2 & = M_1\cos\frac{\theta}{2} -  M_3\sin\frac{\theta}{2}
\end{align}
\end{subequations}
for a negative parity $\Xi$. Here, $\theta$ is the scattering angle, i.e.,
$\cos\theta \equiv \nunit[p] \cdot \nunit[p]'$.

Likewise, the production amplitude $\hat M$ for a $\Xi$ with $J^P=\frac32^\pm$
determined by Eq.~(\ref{eq:Kbar-32-a}) is completely described by four complex
amplitudes given as
\begin{subequations}\label{M_3hf_hel}
\begin{align}
\mathcal{H}_1 & \equiv
\braket{\lambda_\Xi^{} = \tfrac32 | \hat{M} | \lambda_N^{} = \tfrac12}
\nonumber\\
&=
 \pi_\Xi^{} \braket{\lambda_\Xi^{} = -\tfrac32 | \hat{M} | \lambda_N^{}
= -\tfrac12}  ~,
 \\[1ex]
\mathcal{H}_2 & \equiv
\braket{\lambda_\Xi^{} = \tfrac32 | \hat{M} | \lambda_N^{} = -\tfrac12}
\nonumber\\
&= -\pi_\Xi^{} \braket{\lambda_\Xi^{} = -\tfrac32 | \hat{M} |\lambda_N^{}
= \tfrac12}  ~,
 \\[1ex]
\mathcal{H}_3 & \equiv
\braket{\lambda_\Xi^{} = \tfrac12 | \hat{M} | \lambda_N^{} = \tfrac12}
\nonumber\\
&= -\pi_\Xi^{} \braket{\lambda_\Xi^{} = -\tfrac12 | \hat{M} | \lambda_N^{}
= -\tfrac12}  ~,
 \\[1ex]
\mathcal{H}_4& \equiv
\braket{\lambda_\Xi^{} = \tfrac12 | \hat{M} | \lambda_N^{} = -\tfrac12}
\nonumber\\
&= \pi_\Xi^{} \braket{ \lambda_\Xi^{} = -\tfrac12 | \hat{M} | \lambda_N^{}
= \tfrac12} ~.
\end{align}
\end{subequations}
These helicity amplitudes are related to the coefficient functions in
Eq.~(\ref{eq:Kbar-32-a}) by
\begin{subequations}\label{eq:HF-32+}
\begin{align}
\mathcal{H}_1 & =\frac1{\sqrt2} \bigg[ i \cos\frac\theta2 F_1
- \cos\theta \sin\frac\theta2 F_2
\nonumber \\
&\qquad\mbox{}
  -\cos\frac{3\theta}{2} F_3
+ \sin\theta\cos\frac\theta2 F_4\bigg]   ~,
 \\
\mathcal{H}_2 & =\frac1{\sqrt2} \bigg[ i \sin\frac\theta2 F_1
- \cos\theta\cos\frac\theta2 F_2
\nonumber \\
&\qquad\mbox{}
 +\sin\frac{3\theta}{2} F_3 - \sin\theta \sin\frac\theta2 F_4\bigg]   ~,
 \\
\mathcal{H}_3 & = \frac1{\sqrt6} \bigg[ -i \sin\frac\theta2 F_1
+ \left(2-3\cos\theta\right) \cos\frac\theta2 F_2
\nonumber \\
&\qquad\mbox{}
 +3\sin\frac{3\theta}{2} F_3
- \left(1-3\cos\theta\right) \cos\frac\theta2 F_4\bigg]  ~,
 \\
\mathcal{H}_4 & = \frac1{\sqrt6} \bigg[ i \cos\frac\theta2 F_1
+ \left(2+3\cos\theta\right) \sin\frac\theta2 F_2
\nonumber \\
&\qquad\mbox{}
 +3\cos\frac{3\theta}{2} F_3
 - \left(1+3\cos\theta\right) \sin\frac\theta2 F_4\bigg]  ~,
\end{align}
\end{subequations}
for a positive-parity $\Xi$ and by
\begin{subequations}\label{eq:HG-32-}
\begin{align}
\mathcal{H}_1 & = \frac1{\sqrt2}
\bigg[ i \left(2-\cos\theta\right) \cos\frac\theta2 G_1
+ 2 i \sin^3\frac\theta2 G_2
\nonumber \\
&\qquad\mbox{}
 -\cos\theta\cos\frac\theta2 G_3
+ \sin\theta\cos\frac\theta2 G_4\bigg]  ~,
 \\
\mathcal{H}_2 & = \frac1{\sqrt2}
\bigg[ -i \left(2+\cos\theta\right) \sin\frac\theta2 G_1
+ 2 i \cos^3\frac\theta2 G_2
\nonumber \\
&\qquad\mbox{}
 -\cos\theta\sin\frac\theta2 G_3
+\sin\theta\sin\frac\theta2 G_4 \bigg]  ~,
 \\
\mathcal{H}_3 & =\frac1{\sqrt6}
\bigg[ 3i \cos\theta\sin\frac\theta2 G_1
+ 3i \sin\theta\sin\frac\theta2 G_2
\nonumber \\
&\qquad\mbox{}
 + \left(2+3\cos\theta\right) \sin\frac\theta2 G_3
+ (3\cos\theta-1) \cos\frac\theta2 G_4\bigg]  ~,
 \\
\mathcal{H}_4 & = \frac1{\sqrt6}
\bigg[ -3i \cos\theta\cos\frac\theta2 G_1
-3i \sin\theta\cos\frac\theta2 G_2
\nonumber \\
&\qquad\mbox{}
+\left(2-3\cos\theta\right) \cos\frac\theta2 G_3
+\left(3\cos\theta+1\right) \sin\frac\theta2 G_4\bigg]
\end{align}
\end{subequations}
for a negative-parity $\Xi$.

The SDM elements are defined by
\begin{eqnarray}
\rho^{\Xi,i}_{\lambda\lambda'} \equiv
\braket{\lambda | \hat{\rho}^{N,i} |\lambda'}
= \frac12 \braket{\lambda | \hat{M} \sigma_i^{} \hat{M}^\dagger |\lambda'} ,
\label{eq:SDM-elem}
\end{eqnarray}
for $i = 0, \dots, 3$, where $\lambda$ and $\lambda'$ stand for the helicity of
the produced $\Xi$ baryon and $\sigma_0^{} = \mathbb{1}$ is the $2 \times 2$ unit
matrix. For completeness, a relevant part of the SDM formalism for the present
work is presented in Appendix~\ref{app:SDM}. The SDM elements are related by
\begin{subequations}\label{rho_sym_refl_herm}
\begin{align}
\rho^{\Xi,i}_{\lambda,\lambda'} &=(-1)^{i+\lambda-\lambda'}
\rho^{\Xi,i}_{-\lambda,-\lambda'}  ~,
 \\
\rho^{\Xi,i}_{\lambda,\lambda'} &=\rho^{\Xi,i*}_{\lambda',\lambda}
\end{align}
\end{subequations}
due to the symmetry of the spin matrix element~(\ref{mirror_sym1}) and the
hermiticity of Eq.~(\ref{eq:SDM-elem}).

We now relate the SDM elements, $\rho^{\Xi,i}_{\lambda,\lambda'}$, to the
helicity amplitudes $\mathcal{H}_j$ given by Eqs.~(\ref{M_1hf_hel}) and
(\ref{M_3hf_hel}) which determine the reaction amplitudes. The purpose is to
find a set of SDM elements that fixes those helicity amplitudes completely.

\subsection{\boldmath $\Xi$ of $J^P=\frac12^\pm$}

Starting with $J=\frac12$, there are sixteen possible SDM elements
$\rho^{\Xi,i}_{\lambda\lambda'}$. However, only four of them are independent
for a given parity and they determine the amplitudes $\mathcal{H}_1$ and
$\mathcal{H}_2$ apart from an irrelevant overall phase. Inserting
Eq.~(\ref{M_1hf_hel}) into Eq.~(\ref{eq:SDM-elem}), a set of four independent
SDM elements can be determined as
\begin{subequations}\label{eq:HSDM-12}
\begin{align}
 2\rho^0_{\frac12,\frac12} = 2i\pi_\Xi^{} \, \rho^2_{\frac12,-\frac12}
& = |\mathcal{H}_1|^2+|\mathcal{H}_2|^2  ~,
 \\
2\rho^3_{\frac12,\frac12} = 2\pi_\Xi^{} \, \rho^1_{\frac12,-\frac12}
& = |\mathcal{H}_1|^2-|\mathcal{H}_2|^2  ~,
 \\
\rho^2_{\frac12,\frac12} = i\pi_\Xi^{} \, \rho^0_{\frac12,-\frac12}
& =\mbox{Im} \left[ \mathcal{H}_1\mathcal{H}^*_2 \right] ~,
 \\
\rho^1_{\frac12,\frac12} = -\pi_\Xi^{} \, \rho^3_{\frac12,-\frac12}
& = \mbox{Re} \left[ \mathcal{H}_1\mathcal{H}^*_2 \right]  ~,
\end{align}
\end{subequations}
where the superindex $\Xi$ in $\rho^{\Xi,i}_{\lambda\lambda'}$ was dropped for
simplicity. A complete list of SDM elements $\rho^i_{\lambda,\lambda'}$ in
terms of helicity amplitudes $\mathcal{H}_i$ is given in
Appendix~\ref{app:SDM_form}.

The SDM elements are directly related to the observables defined by
Eq.~(\ref{eq:Kbar-12-b-alt}). For example, from Eqs.~(\ref{rho_sym_refl_herm})
and (\ref{rho_norm}), we have
\begin{subequations}\label{eq:SOSDM-12}
\begin{align}
 \frac{d\sigma}{d\Omega}  &= 2\rho^0_{\frac12,\frac12}  ~,
 &
 \frac{d\sigma}{d\Omega} K_{yy'} &= 2i\rho^2_{\frac12,-\frac12}  ~,
 \\
 \frac{d\sigma}{d\Omega} T_y  &= 2 \rho^2_{\frac12,\frac12} ~,
&
\frac{d\sigma}{d\Omega} P_{y'}    &= 2i \rho^0_{\frac12,-\frac12}  ~,
 \\
 \frac{d\sigma}{d\Omega} K_{xx'}  &=  2 {\rho}^1_{\frac12,-\frac12} ~,
&
\frac{d\sigma}{d\Omega} K_{zz'}    &= 2\rho^3_{\frac12,\frac12}  ~,
 \\
 \frac{d\sigma}{d\Omega} K_{xz'} &= 2 \rho^1_{\frac12,\frac12} ~,
&
\frac{d\sigma}{d\Omega} K_{zx'}   &=  2 \rho^3_{\frac12,-\frac12}  ~,
\end{align}
\end{subequations}
where the primed Cartesian components correspond to the rotated frame (see
Fig.~\ref{fig:1}; note that $y'\equiv y$).

 From Eqs.~(\ref{eq:Kbar-12-e}) and (\ref{eq:SOSDM-12}), we see, in
particular, that
\begin{equation}
K_{yy'} = \frac{i\rho^2_{\frac12, -\frac12}}{\rho^0_{\frac12, \frac12}}= \pi_\Xi^{} ~.
\label{eq:KyySDM-12a}
\end{equation}
More generally, in terms of the SDM elements, one obtains
\begin{equation}
 (-1)^{\frac 12-\lambda'} \frac{i\rho^2_{\lambda,-\lambda'}}
{\rho^0_{\lambda,\lambda'}} =
(-1)^{\frac 12-\lambda'} \frac{\rho^1_{\lambda,-\lambda'}}
{\rho^3_{\lambda,\lambda'}} =\pi_\Xi^{} ~.
\label{eq:KyySDM-12}
\end{equation}
This result reveals that one needs to measure two SDM elements to determine the
parity of the $\Xi$ baryon: either $\rho^0_{\lambda,\lambda'}$ with unpolarized
target nucleon and $\rho^2_{\lambda,-\lambda'}$ with polarized target nucleon
along the direction $\nunit_2^{}\equiv\nunit'_2$ perpendicular to the reaction
plane, or $\rho^1_{\lambda,\lambda'}$ with transversally polarized target along
$\nunit_1^{}$, and $\rho^3_{\lambda,-\lambda'}$ with longitudinally polarized
target along $\nunit_3^{} \equiv \nunit[p]$.
Note that $\rho^0_{\lambda,\lambda'}$
is directly related to the cross section $d\sigma/d\Omega$ when
$\lambda=\lambda'$.

\subsection{\boldmath $\Xi$ of $J^P = \frac32^\pm$}

For $J=3/2$, analogously to the $J=1/2$ case, inserting Eq.~(\ref{M_3hf_hel})
into Eq.~(\ref{eq:SDM-elem}), the SDM elements are related to the four
amplitudes $\mathcal{H}_i$ $(i=1,\ldots,4)$ by
\begin{subequations}
\def\theparentequation{\arabic{parentequation}}
\begin{align}
2 \rho^0_{\frac32,\frac32} & = |\mathcal{H}_1|^2 + |\mathcal{H}_2|^2  ~,
 \\
2 \rho^1_{\frac32,-\frac32} & =
\pi_\Xi^{} \left( |\mathcal{H}_1|^2 - |\mathcal{H}_2|^2 \right)  ~,
 \\
-i \rho^0_{\frac32,-\frac32} & =
\pi_\Xi^{} \, \mbox{Im} \left[ \mathcal{H}_2 \mathcal{H}_1^* \right]  ~,
 \\
\rho^1_{\frac32,\frac32} & =
\mbox{Re} \left[ \mathcal{H}_2 \mathcal{H}_1^* \right]  ~,
\\[1ex]\stepcounter{parentequation}\setcounter{equation}{0}
2\rho^0_{\frac12,\frac12} & = |\mathcal{H}_3|^2 + |\mathcal{H}_4|^2  ~,
 \\
2\rho^1_{\frac12,-\frac12} & =
\pi_\Xi^{} \left(|\mathcal{H}_4|^2 - |\mathcal{H}_3|^2\right)  ~,
 \\
-i\rho^0_{\frac12,-\frac12} & =
\pi_\Xi^{} \, \mbox{Im} \left[ \mathcal{H}_3 \mathcal{H}_4^* \right] ~,
 \\
\rho^1_{\frac12,\frac12} & =
\mbox{Re} \left[ \mathcal{H}_3 \mathcal{H}_4^* \right]  ~,
\\[1ex]\stepcounter{parentequation}\setcounter{equation}{0}
2\rho^1_{\frac32,-\frac12} & =
\pi_\Xi^{} \left( \mathcal{H}_2 \mathcal{H}_4^*
- \mathcal{H}_1 \mathcal{H}_3^* \right)  ~,
 \\
2\rho^0_{\frac32,\frac12} & =
\mathcal{H}_2 \mathcal{H}_4^* + \mathcal{H}_1 \mathcal{H}_3^*  ~,
\\[1ex]\stepcounter{parentequation}\setcounter{equation}{0}
2\rho^0_{\frac32,-\frac12} & =
\pi_\Xi^{} \left( \mathcal{H}_1 \mathcal{H}_4^*
- \mathcal{H}_2 \mathcal{H}_3^* \right)  ~,
 \\
2\rho^1_{\frac32,\frac12} & =
\mathcal{H}_1 \mathcal{H}_4^*
+ \mathcal{H}_2 \mathcal{H}_3^*  ~.
\end{align}
\end{subequations}
A complete list of SDM elements $\rho^{\Xi,i}_{\lambda,\lambda'}$ in terms of
the helicity amplitudes $\mathcal{H}_i$ is given in
Appendix~\ref{app:SDM_form}.

As mentioned in the previous section, a set of eight independent SDM elements
(e.g., $\rho^0_{\frac12,\frac12}$, $\rho^0_{\frac32,\frac32}$,
$\mathrm{Re}[\rho^0_{\frac32,\frac12}]$, $\mathrm{Im}[\rho^0_{\frac32,\frac12}]$,
$\mathrm{Re}[\rho^0_{\frac32,-\frac12}]$, $\rho^1_{\frac12,-\frac12}$,
$\rho^1_{\frac32,-\frac32}$, $\mathrm{Re}[\rho^1_{\frac32,\frac12}]$)
determines all four
helicity amplitudes ${\cal H}_i,\ (i=1,\ldots,4)$, apart from an irrelevant
overall phase. The analysis here is analogous to the one carried out in
Ref.~\cite{CT97} for pion photoproduction.

Equations (\ref{rho_sym_refl_herm}) and (\ref{rho32_obs}) can be used to
 calculate $K_{yy}$ in terms of SDM elements involving $\Xi$ of spin-3/2.
This leads to
\begin{equation}
K_{yy} =
\frac{i\left(\rho^2_{\frac32,-\frac32}-\rho^2_{\frac12,-\frac12}\right)}
{\rho^0_{\frac32,\frac32}+\rho^0_{\frac12,\frac12}} = \pi_\Xi^{}
\label{eq:KyySDM-32}
\end{equation}
or, more generally,
\begin{equation}
(-1)^{\frac 32-\lambda'} \,
\frac{i \rho^2_{\lambda,-\lambda'}}{\rho^0_{\lambda,\lambda'}} =
(-1)^{\frac 32-\lambda'}\,
\frac{\rho^1_{\lambda,-\lambda'}}{\rho^3_{\lambda,\lambda'}} =
\pi_\Xi^{} ~.
\label{SDM32_prty}
\end{equation}

Equations (\ref{eq:KyySDM-12a}), (\ref{eq:KyySDM-12}), (\ref{eq:KyySDM-32}),
and (\ref{SDM32_prty}) can be extended to an arbitrary spin $J$ of the $\Xi$
baryon, i.e.,
\begin{align}
K_{yy}& =\frac{i\sum_\lambda (-1)^{J-\lambda} \rho^2_{\lambda,-\lambda}}
{\sum_\lambda \rho^0_{\lambda,\lambda}} =\pi_\Xi^{}
\label{eq:KyySDM-J}
\end{align}
and
\begin{equation}
(-1)^{J-\lambda'} \,
\frac{i\rho^2_{\lambda,-\lambda'}}{\rho^0_{\lambda,\lambda'}} =
(-1)^{J-\lambda'} \,
\frac{\rho^1_{\lambda,-\lambda'}}{\rho^3_{\lambda,\lambda'}} =\pi_\Xi^{} ~.
\label{SDMJ_prty}
\end{equation}
Note here that in the last expression two SDM elements are sufficient to
determine the parity, whereas one needs a whole sum of SDM elements to achieve
the same in terms of $K_{yy}$.

\subsection{Extracting SDM elements from experiment}

To extract SDM elements from experiment, following Chung~\cite{Chung71} and
Biagi \textit{et al.}~\cite{BBBB87b}, one may relate them to moments, $H^i$, and weak
decay-asymmetry parameters. Their definitions and further full details are
given in Appendix~\ref{app:measure_SDM}.
Here, we only present some pertinent results.

For a spin-$J$ $\Xi$ undergoing a single weak decay process, one obtains the
SDM element
\begin{equation}
\rho^{\Xi,i}_{\lambda^{}_\Xi,\lambda'^{}_\Xi} = \sum_{L} \frac{2L+1}{2J+1}
\braket{J\, \lambda_\Xi' \, L \, M | J\, \lambda_\Xi^{} } \,
t^{J,i}_{LM}  ~,
\label{rho_ta}
\end{equation}
where $M=\lambda^{}_\Xi-\lambda'_\Xi$.
The coefficients $t^{J,i}_{LM}$ can be determined from the ratio of the
moments,
\begin{equation}
\left(\frac{d\sigma}{d\Omega}\right) \frac{H^i(L,M)}{H^0(0,0)} =  \zeta_L^{} \,
t^{J,i}_{LM} \, \braket{ J\, \tfrac12 \, L \, 0 | J \, \tfrac12 }~,
\label{HLM_1hfa}
\end{equation}
where $\zeta_L^{} = 1$ for even $L$ and $\zeta_L^{} = \alpha_\Xi^{}$ for odd $L$, with
$\alpha_\Xi^{}$ denoting the $\Xi$ decay-asymmetry parameter.

Note that  since all moments vanish identically for $L>2J$,
Eq.~(\ref{HLM_1hfa}) offers  a way of determining the spin of the $\Xi$
undergoing a single (weak) decay by measuring  the moments as a function of
$L$. In other words, the nonvanishing $H^i(L,M)$ with the largest $L$ value for
some $i$ and $M$ determines $J$ as $J=L/2$. Experimentally, of course, this
may be challenging since it is not a priori clear how small the measured
values of the next higher moment $H^i(L+1,M)$ would need to be for being compatible
with zero. And, moreover, one would need to confirm that the smallness of this
moment is not accidental.

Similar results are obtained for excited $\Xi$ resonances, $\Xi^*$,
undergoing a double decay process, as discussed in
Appendix~\ref{app:measure_SDM}.
In this case, we have,
\begin{subequations}\label{HLM_3hfa}
\begin{align}
\left(\frac{d\sigma}{d\Omega} \right) \frac{H^i(0,0,L,M)}{H^0(0,0,0,0)} &=  t^{J,i}_{LM}
\braket{ J \, \tfrac12 \, L \, 0 | J \, \tfrac12 }
\\
\intertext{for even $L$ and}
\left(\frac{d\sigma}{d\Omega} \right) \frac{H^i(1,0,L,M)}{H^0(0,0,0,0)} &=
\frac{\alpha^{}_\Lambda}{3}\, t^{J,i}_{LM} \,
\braket{ J \, \tfrac12 \, L \, 0 | J \, \tfrac12 }
\end{align}
\end{subequations}
for odd $L$. Here, $\alpha_\Lambda^{}$ denotes the $\Lambda$ decay-asymmetry
parameter for the decay chain $\Xi^* \to \Lambda + \bar{K}$ followed by $\Lambda
\to N + \pi$. In the case of  $\Xi^* \to \Xi + \pi$ followed by $\Xi \to
\Lambda + \pi$ instead, $\alpha^{}_\Lambda$ needs to be replaced by
$\alpha^{}_\Xi$.

For $\Xi$ resonances decaying along the double decay chain specified in
Eq.~(\ref{eq:Xi-dd}), the result of the corresponding moments given by
Eq.~(\ref{HlmLM}) leads to~\cite{BBBB87b}
\begin{equation}
\frac{H^0(1,\pm1,L,M)}{H^0(1,0,L,M)} = \pi_\Xi^{} (-1)^{J+\frac12}
\frac{2 J+1}{\sqrt{2L(L+1)}}
\label{eq:SJ-determ}
\end{equation}
for an unpolarized target and for odd values of $L (\leq 2J)$. This offers a
way of determining the spin and parity of the excited $\Xi$ resonance
simultaneously.

\section{Summary}
\label{sec:summary}

A model-independent analysis of the $\bar{K} + N \to K + \Xi$ reaction has been
performed. Following the method of Ref.~\cite{NL05}, we derived the most
general spin structure of the reaction amplitude, consistent with basic
symmetries, for $\Xi$ baryons of $J^P=\frac12^\pm$ and $\frac 32^\pm$. The
coefficients multiplying each spin structure have been presented in
partial-wave-decomposed form, thus permitting partial-wave
analyses, once sufficient data become available for these reactions. The method of
Ref.~\cite{NL05} is general, and can be applied, in principle, to derive the
structure of the reaction amplitude involving higher-spin $\Xi$ production.

Furthermore, a minimal set of independent observables required to determine
completely the reaction amplitude has been identified. In addition to the
unpolarized cross sections, one also needs single- and double-spin observables,
which poses a formidable experimental challenge, in particular, since one needs
to measure the polarization of the outgoing $\Xi$. Note that, for the $\Xi$ of
spin-1/2, there are two complex amplitudes to be determined, whereas for the
$\Xi$ of spin-3/2, there are four complex amplitudes. We then formulated the
problem using the SDM approach and expressed the spin observables in terms of
the SDM elements. Following Ref.~\cite{BBBB87b, Chung71}, it was shown that the latter
can be extracted from  the moments associated with the $\Xi$ decay processes in
conjunction with the self-analyzing nature of the hyperon ($\Lambda$ or $\Xi$)
resulting from the subsequent decay of the $\Xi$ produced in the primary
reaction. The moments, in turn, can be extracted from the measurement of the
angular distribution of the decay products.

Since the determination of the spin and parity quantum numbers is a fundamental
part of any spectroscopy study, reflection symmetry about
 the reaction plane has been exploited, in particular, to
show that, apart from the spin-transfer coefficient $K_{yy}$, the ratio of the
SDM elements given by Eq.~(\ref{SDMJ_prty}) determines the parity of a $\Xi$
resonance with an arbitrary spin. Furthermore, the moments given by
Eq.~(\ref{eq:SJ-determ}) determine the spin and parity of the $\Xi$ resonance
simultaneously~\cite{BBBB87b}.

We also mention that the present analysis applies as given only to $\Xi$
resonances that are sufficiently narrow to permit them being treated like
on-shell particles. For broad resonances, a partial-wave analysis would be
required to extract them from experimental data.

In summary, the present analysis provides the model-independent framework for
developing reliable reaction theories of $\Xi$ production to help in the
planning of future experimental efforts in $\Xi$ baryon spectroscopy.
This will also help in analyzing the data to understand the production
mechanisms of $\Xi$ baryons.

\acknowledgments

We thank Johann Haidenbauer for many useful discussions and  for a careful
reading of the manuscript.
Also, we thank Lei Guo for help in clarifying some of the experimental procedures.
We are also grateful to the Asia Pacific Center for
Theoretical Physics for their warm hospitality during the topical research
program. This work was supported by the National Research Foundation of Korea
funded by the Korean Government (Grant No.\ NRF-2011-220-C00011). The work of
K.N. was also supported in part by the FFE-COSY Grant No.\ 41788390.

\appendix

\section{Partial-Wave Decomposition}\label{app:A}

In this Appendix, we give the partial wave decomposition of the coefficients
multiplying each spin structure of the reaction amplitudes in
Eqs.~(\ref{eq:Kbar-12-a}) and (\ref{eq:Kbar-32-a}) for $\Xi$ with spin-parity
$J^P=\frac12^\pm$ and $\frac 32^\pm$. The partial-wave expansion of the
(plane-wave) matrix element  $\hat{M}$ in Eqs.~(\ref{eq:Kbar-12-a})
and (\ref{eq:Kbar-32-a}) is
\begin{align}
& \braket{S' M_{S'} | \hat M (\bm{p}', \bm{p}) | S M_S} =
\nonumber \\
& \quad \sum i^{L-L'}
\braket{ S \, M_S \, L \, M_L | J \, M_J}
\braket{ S' \, M_{S'} \, L' \, M_{L'} | J \, M_J }
\nonumber \\
& \qquad \times M^{TJ}_{L'L}(p',p) Y_{L'M_{L'}}(\hat{\bm{p}}')
Y^*_{LM_L}(\hat{\bm{p}})\, \hat{P}_T ~,
\label{eq:A1}
\end{align}
where $S, L, J, T$ stand for the total spin, orbital angular momentum, total
angular momentum, and the total isospin, respectively, of the initial
$\bar{K}N$ state. The corresponding projection quantum numbers are denoted by
$M_S$, $M_L$, and $M_J$. The primed quantities represent the corresponding
quantum numbers of the final $K\Xi$ state.  The summation runs over all quantum
numbers not specified in the left-hand side of Eq.~(\ref{eq:A1}).
The relative momenta of the initial $\bar{K}N$ and final $K\Xi$ states are
denoted by $\bm{p}$ and $\bm{p}'$, respectively, and $p=|\bm{p}|$,
$p'=|\bm{p}'|$. In the following, without loss of generality, we choose
$\nunit_3^{}$ along the momentum $\bm{p}$ of the nucleon in the
CM system, i.e., $\nunit_3^{} \equiv \nunit[p]$ as specified in
Fig.~\ref{fig:1}.
In Eq.~(\ref{eq:A1}), $\hat{P}_T$ stands for the total isospin projection operator
onto the isospin singlet ($T=0$) and isospin triplet ($T=1$) states,
\begin{equation}
\hat{P}_0  = \frac14 (1-\bm{\tau}^{}_1\cdot \bm{\tau}^{}_2)
 \quad\text{and}\quad
\hat{P}_1  = \frac14 (3+\bm{\tau^{}}_1\cdot \bm{\tau}^{}_2)~,
\label{eq:isoproj}
\end{equation}
where the $\bm{\tau}^{}_i$ ($i=1,2$) are the usual vectors made out of isospin
Pauli matrices.

For a $\Xi$ of $J^P=\frac12^\pm$, following Ref.~\cite{NL05}, the coefficients
$M_i$ in Eq.~(\ref{eq:Kbar-12-a}) are given by
\begin{widetext}
\begin{subequations}\label{eq:A2}
\begin{align}
M_0 & =  \frac{1}{4\pi} \sum_{L',T} \left[ (L'+1) \, M^{TJ_+}_{L'L'}(p', p)
+ L'\, M^{TJ_-}_{L'L'}(p', p)\right]  P_{L'}(\hat{\bm{p}} \cdot \hat{\bm{p}}') \, \hat{P}_T ~,
 \\
M_2 & =  \frac{i}{4\pi} \sum_{L',T} \left[ M^{TJ_-}_{L'L'}(p', p)
- M^{TJ_+}_{L'L'}(p', p) \right] P^1_{L'}(\hat{\bm{p}} \cdot \hat{\bm{p}}') \, \hat{P}_T ~,
 \\
M_1 & =  \frac{i}{4\pi} \sum_{L',T} \left[ M^{TJ_-}_{L'\,L'-1}(p', p)
+ M^{TJ_+}_{L'\,L'+1}(p', p) \right]  P^1_{L'}(\hat{\bm{p}} \cdot \hat{\bm{p}}') \,  \hat{P}_T ~,
 \\
M_3 & =  \frac{i}{4\pi} \sum_{L',T} \left[ (L'+1) \, M^{TJ_+}_{L'\,L'+1}(p', p)
 - L' \,  M^{TJ_-}_{L'\,L'-1}(p', p)\right] P_{L'}(\hat{\bm{p}} \cdot \hat{\bm{p}}') \, \hat{P}_T ~,
\end{align}
\end{subequations}
where $J_\pm \equiv L'\pm\frac12$, and $P_{L'}(x)$ and $P^1_{L'}(x)$ denote
the Legendre and associated Legendre functions, respectively.%
\footnote{Here, the phase convention for the associated Legendre function is
such that $P^1_{1}(x) = + \sin(x)$.}
The amplitudes $M_i$ here are operators in isospin space whose actions are
specified by the projectors $\hat{P}_T$ defined in  Eq.~(\ref{eq:isoproj}).

Likewise, for a $\Xi$ of $J^P=\frac32^\pm$, following Ref.~\cite{NL05}, the
coefficients $F_i$ and $G_i$ in Eq.~(\ref{eq:Kbar-32-a}) are given by
\begin{subequations}
\begin{align}
F_1 & =  i \frac{3}{8\pi} \sum_{J, L', T} (-1)^{L'+J+\frac{3}{2}}[J]^2
\frac{[L']}{\sqrt{L'(L'+1)}}
\begin{Bmatrix}
  \frac{1}{2} & L' & J \\   L' & \frac{3}{2} & 1
  \end{Bmatrix}
M^{JT}_{L'L'}(p',p)
P_{L'}^1(\hat{\bm{p}}' \cdot \hat{\bm{p}}) \, \hat{P}_T ~,
 \\
F_2 & = \frac{1}{\sqrt{2}} \sum_{J, L', L, T} i^{L-L'} (-1)^{J+\frac{1}{2}} [J]^2
 \begin{Bmatrix}
  \frac{1}{2} & L & J \\   L' & \frac{3}{2} & 2
  \end{Bmatrix}
M^{JT}_{L'L}(p',p)  \, a_{L'L}^{} \, \hat{P}_T ~,
 \\
F_3 & = \frac{1}{2\sqrt{2}} \sum_{J, L', L, T} i^{L-L'} (-1)^{J+\frac{1}{2}} [J]^2
\begin{Bmatrix}
  \frac{1}{2} & L & J \\   L' & \frac{3}{2} & 2
  \end{Bmatrix}
M^{JT}_{L'L}(p',p) \, b_{L'L}^{} \, \hat{P}_T ~,
 \\
F_4 & = \frac{1}{\sqrt{2}} \sum_{J, L', L, T} i^{L-L'} (-1)^{J+\frac{1}{2}} [J]^2
\begin{Bmatrix}
  \frac{1}{2} & L & J \\   L' & \frac{3}{2} & 2
  \end{Bmatrix}
M^{JT}_{L'L}(p',p)  \, c_{L'L}^{} \, \hat{P}_T ~,
\end{align}
\end{subequations}
 and
\begin{subequations}\label{eq:A3}
\begin{align}
G_1 & = \frac{1}{2\sqrt{2}} \sum_{J, L', L, T} i^{L-L'} (-1)^{J+\frac{1}{2}} [J]^2
\begin{Bmatrix}
  \frac{1}{2} & L & J \\   L' & \frac{3}{2} & 2
  \end{Bmatrix}
M^{JT}_{L'L}(p',p)  \, a'_{L'L} \, \hat{P}_T ~,
 \\
G_2 & = \frac{1}{2\sqrt{2}} \sum_{J, L', L, T} i^{L-L'} (-1)^{J+\frac{1}{2}} [J]^2
\begin{Bmatrix}
  \frac{1}{2} & L & J \\   L' & \frac{3}{2} & 2
  \end{Bmatrix}
M^{JT}_{L'L}(p',p)  \, b'_{L'L}  \, \hat{P}_T ~,
 \\
G_3 & =  \sqrt{\frac{3}{2}}\frac{1}{4\pi} \sum_{J, L', L, T} i^{L-L'}(-1)^{J+\frac{1}{2}}[J]^2
\frac{[LL']}{\sqrt{L'(L'+1)}}
\braket{ L \, 0 \, L' 1\, | 1 \, 1 }
  \begin{Bmatrix}
  \frac{1}{2} & L & J \\   L' & \frac{3}{2} & 1
  \end{Bmatrix}
M^{JT}_{L'L}(p',p)
P_{L'}^1(\hat{\bm{p}}' \cdot \hat{\bm{p}}) \, \hat{P}_T ~,
 \\
G_4 & = \frac{\sqrt{3}}{8\pi} \sum_{J, L', L, T} i^{L-L'} (-1)^{J+\frac{3}{2}} [J]^2 [LL']
\braket{ L \, 0 \, L' \, 0 | 1 \, 0 }
\begin{Bmatrix}
  \frac{1}{2} & L & J \\   L' & \frac{3}{2} & 1
  \end{Bmatrix}
M^{JT}_{L'L}(p',p) P_{L'}(\hat{\bm{p}}' \cdot \hat{\bm{p}}) \, \hat{P}_T  ~,
\end{align}
\end{subequations}
where we introduced the notation $[J] \equiv \sqrt{2J+1}$ and $[j_1  j_2]
\equiv [j_1] \, [j_2]$. The summations extend over all the quantum numbers $J,
L', L$ and $T$. Note that total parity conservation imposes the
condition $(-1)^{L'+L} = \pm 1$ as the parity of the $\Xi$ baryon is
$\pi_\Xi^{} =\pm1$. The coefficients $a_{L'L}^{}$,  $b_{L'L}^{}$, etc, are
given by
\begin{subequations}
\begin{align}
a_{L',L}^{} &= 2\frac{[LL']}{4\pi} \langle L \, 0 \, L' \, 2 \mid 2 \, 2 \rangle \,
\sqrt{\frac{(L'-2)!}{(L'+2)!}} \, P_{L'}^2(\hat{\bm{p}}' \cdot \hat{\bm{p}}) ~,
 \\
b_{L',L}^{} &=2\frac{[LL']}{4\pi}
\langle L \, 0 \, L' \, 1 \mid 2 \, 1 \rangle \, \sqrt{\frac{(L'-1)!}{(L'+1)!}}
\, P_{L'}^1(\hat{\bm{p}}' \cdot \hat{\bm{p}})  ~,
 \\
c_{L',L}^{} &=\frac{[LL']}{4\pi}\bigg[
\langle L \, 0 \, L' \, 2 \mid 2 \, 2 \rangle \,
\sqrt{\frac{(L'-2)!}{(L'+2)!}} \, P_{L'}^2(\hat{\bm{p}}' \cdot \hat{\bm{p}})
 +\sqrt{\frac32} \langle L \, 0 \, L' \, 0 \mid 2 \, 0 \rangle \,
P_{L'}(\hat{\bm{p}}' \cdot \hat{\bm{p}})  \bigg]  ~,
 \\
a'_{L',L} &= i b_{L',L}^{} ~,
 \\
b'_{L',L} &= -i a_{L',L}^{} ~.
\end{align}
\end{subequations}
\end{widetext}

\section{SDM formalism}\label{app:SDM}

A density operator can be used to describe an
ensemble of quantum states. It is defined as
\begin{equation}
\hat\rho \equiv \sum_\psi I_\psi\ket{\psi}\bra{\psi} \ ,
\label{rho_def}
\end{equation}
where $I_\psi$
denotes the probability of finding an element of the ensemble in the state
$\psi$, subject to the condition $\sum_\psi I_\psi =1$.
For the present application, the states $\ket{\psi}$ are the spin
states of the initial state, $N$, or the final state, $\Xi$.
For the initial nucleon state, the spin-density operator reads
\begin{align}
\hat\rho \to \hat\rho^N & \equiv
\sum_{\psi_N^{}} I_{\psi_N^{}} \ket{\psi_N^{}} \bra{\psi_N^{}}
\nonumber \\
& =\frac 12 \left( \mathbb{1} + \bm{P} \cdot \bm{\sigma} \right)  ~,
\label{rhoN_def}
\end{align}
where $\bm{\sigma}=(\sigma_1^{},\sigma_2^{},\sigma_3^{})$ denotes the vector
formed of Pauli spin matrices and $\bm{P}$ is the polarization vector of the
nucleon which is the difference between the probability of finding the nucleon
in the $m_N=+\frac 12$ spin state and the probability of finding the nucleon in
the $m_N=-\frac 12$ state ($m_N$ is the spin projection quantum number along
the $\bm{P}$ direction) or symbolically $|I_{\psi_+}-I_{\psi_-}|=|\bm{P}|$.

An unpolarized ensemble has $\bm{P}=0$. The trace of this spin-$\frac 12$
density matrix is normalized to 1. By introducing the notation $\sigma_0^{}
\equiv \mathbb{1}$ and $P_{0} \equiv 1$, the nucleon SDM in
Eq.~(\ref{rhoN_def}) can be rewritten as
\begin{equation}
\hat\rho^N =  \frac{\mathbb{1}+\bm{P} \cdot \bm{\sigma}}{2}
= \sum_{i=0}^3 P_{i}\, \hat{\rho}^{N,i}
\label{rhoN_def1}
\end{equation}
with
\begin{equation}
\hat\rho^{N,i} \equiv \frac12 \sigma_i^{}
\label{rhoN_def2}
\end{equation}
for $i=0, \dots, 3$.

The spin-density operator for a produced $\Xi$ particle, $\hat{\rho}^\Xi$, can
be expressed in terms of the production amplitude $\hat M$ which is an
operator that maps the initial nucleon spin state $\psi_N^{}$ into a spin-state of
the $\Xi$. In the helicity basis
for the produced $\Xi$, the corresponding spin-density matrix elements read
\begin{equation}
\rho^\Xi_{\lambda\lambda'}(\psi_N^{}) \equiv
\bra{\lambda}\hat M\ket{\psi_N^{}} \bra{\psi_N^{}} \hat M^\dagger \ket{\lambda'} ,
\end{equation}
where $\lambda$ and $\lambda'$ enumerate the $\Xi$'s helicities.

When the beam of anti-Kaons scatters off an ensemble of nucleons, one needs
to average over all nucleon spin states with their appropriate probability
weights, i.e.,
\begin{align}
\rho^\Xi_{\lambda,\lambda'} &\equiv
\sum_{\psi_N^{}} I_{\psi_N^{}} \rho^\Xi_{\lambda\lambda'}(\psi_N^{})
\nonumber \\
&= \sum_{\psi_N^{}} I_{\psi_N^{}} \bra{\lambda} \hat M \ket{\psi_N^{}}
\bra{\psi_N^{}} \hat M^\dagger \ket{\lambda'}
\nonumber \\
& = \bra{\lambda} \hat M\hat\rho^N\hat M^\dagger \ket{\lambda'}
\nonumber \\
& = \bra{\lambda} \hat{\rho}^\Xi \ket{\lambda'} ~,
\end{align}
where Eq.~(\ref{rho_def}) was used to show that the $\Xi$ spin-density operator is
given by~\cite{SSW70}
\begin{equation}
\hat\rho^\Xi = \hat M \hat\rho^N \hat M^\dagger  ~.
\label{eq:SDM-XiN}
\end{equation}
Using Eq.~(\ref{rhoN_def1}), we may write
\begin{equation}
\hat\rho^\Xi =\sum_{i=0}^3 P_{i} \, \hat\rho^{\Xi,i}  ~,
\end{equation}
where
\begin{equation}	
\hat\rho^{\Xi,i} \equiv \frac 12 \hat M \sigma_i^{} \hat M^\dagger ~,
\label{rhoXii_def}
\end{equation}
for $i=0, \dots, 3$. Here, $\hat\rho^{\Xi,0}$ and $\hat\rho^{\Xi,j}$
($j=1,2,3$) provide the respective contributions for unpolarized and polarized initial
nucleons.

For a $\Xi$ baryon of spin-1/2, comparing Eqs.~(\ref{eq:Kbar-12-b}) and
(\ref{rhoXii_def}) gives
\begin{subequations}
\begin{align}
\frac{d\sigma}{d\Omega}  & = \frac12\,
\mbox{Tr} \left[ \hat{M}\hat{M}^\dagger \right]
= \mbox{Tr} \left [\hat\rho^{\Xi,0} \right]  ~,
 \\
\frac{d\sigma}{d\Omega} T_i  & = \frac12\,
\mbox{Tr} \left[ \hat{M}\sigma_i^{}\hat{M}^\dagger \right]
= \mbox{Tr} \left[\hat\rho^{\Xi,i}\right]  ~,
 \\
\frac{d\sigma}{d\Omega} P_i &  = \frac12\,
\mbox{Tr} \left[ \hat{M}\hat{M}^\dagger\sigma_i^{} \right]
= \mbox{Tr} \left[\hat\rho^{\Xi,0}\sigma_i^{} \right]  ~,
 \\
\frac{d\sigma}{d\Omega} K_{ij}  & = \frac12\,
\mbox{Tr} \left[ \hat{M} \sigma_i^{} \hat{M}^\dagger \sigma_j^{} \right]
= \mbox{Tr} \left[\hat\rho^{\Xi,i}\sigma_j^{}\right]  ~.
\label{rho_norm}
\end{align}
\end{subequations}

When a $\Xi$ baryon of spin-3/2 (or higher-spin) is involved, there will be many more
possible degrees of polarization than the spin-1/2 case. For the particular
case of the spin-transfer coefficient, $K_{ij}$, discussed in connection to the
parity of $\Xi$, its definition given in  Eq.~(\ref{eq:Kbar-12-b-alt-d}) has to
be generalized. For this purpose, we first introduce the operator
$\Omega(\bm{J}\cdot\nunit)$ as
\begin{equation}
\Omega(\bm{J}\cdot\nunit) \equiv \sum_{M=-J}^{+J} (-1)^{\frac12-M}\;
\textsf{P}^{J,M}_{\nunit}  ,
\end{equation}
where $\textsf{P}^{J,M}_{\nunit}$
denotes the spin-projection operator onto an arbitrary direction $\nunit$ for
an arbitrary half-integer spin $J$. It can be explicitly calculated as
\begin{align}
\textsf{P}^{J,M}_{\nunit}
&=\sideset{}{'}\prod^{+J}_{m=-J}\frac{m-\bm{J} \cdot \nunit}{m-M}~,
\label{eq:arbspinproj}
\end{align}
where the prime indicates that the factor with $m=M$ is omitted. Here, $\bm{J}
\equiv ( J_1, J_2, J_3 )$ stands for the generator of spin-$J$ rotation. This
expression provides a rotationally invariant polynomial of order $2J$ in
$\bm{J} \cdot \nunit$ that is a generalization to arbitrary  spin of the usual
$(1\pm \bm{\sigma}\cdot \nunit)/2$ projectors for spin-1/2.

With the spin-projection operator defined above, the spin-transfer coefficient
involving a $\Xi$ baryon with an arbitrary spin $J$ is now generalized to
\begin{equation}
\frac{d\sigma}{d\Omega}K_{ba}  =
\frac12\, \mbox{Tr} \left[ \hat{M}\, \bm{\sigma}\cdot\nunit[b] \,
\hat{M}^\dagger \,\Omega(\bm{J}\cdot\nunit[a]) \right]~,
\label{rho32_obs}
\end{equation}
where $\nunit[b]$ and $\nunit[a]$ are the spin directions. For $J=1/2$, in view
of $\Omega(\bm{J}\cdot\nunit[a]) \to \bm{\sigma}\cdot \nunit[a]$, this reduces
to the familiar expression (\ref{eq:Kbar-12-b-alt-d}), of course. For Cartesian
directions $\nunit[b]=\nunit_i^{}$ and $\nunit[a]=\nunit'_j$, in particular,
Eq.~(\ref{rho32_obs}) may be written as
\begin{equation}
\frac{d\sigma}{d\Omega}K_{ij'}  =
 \mbox{Tr} \left[ \hat\rho^{\Xi,i}\Omega^J_{j'} \right]  ,
\label{rho32_obs-Cartesian}
\end{equation}
where $\Omega^J_{j'} \equiv \Omega(\bm{J}\cdot\nunit'_j)$.

For the Cartesian frame $\{\nunit'_1,\nunit'_2\equiv\nunit^{}_2,\nunit'_3\equiv
\nunit[p]'\}$ aligned with the momentum $\bm{p}'$ of the outgoing $\Xi$ (see
Fig.~\ref{fig:1}), explicit expressions for $\Omega^J_{j'}$ are found as
\begin{subequations}\label{def_Omega}
\begin{align}
\Omega^{\frac32}_{x'} &=
\begin{pmatrix}
  0 & 0 & 0 & -1  \\   0 & 0 & -1 & 0  \\ 0 & -1 & 0 & 0 \\ -1 & 0 & 0 & 0
  \end{pmatrix}  ~,
 \\
\Omega^{\frac32}_{y'} &=
\begin{pmatrix}
  0 & 0 & 0 & -i  \\   0 & 0 & i & 0  \\ 0 & -i & 0 & 0 \\ i & 0 & 0 & 0
  \end{pmatrix}  ~,
 \\
\Omega^{\frac32}_{z'} &=
\begin{pmatrix}
  -1 & 0 & 0 & 0 \\  0 & 1 & 0 & 0  \\ 0 & 0 & -1 & 0  \\ 0 & 0 & 0 & 1
  \end{pmatrix}   ~,
\end{align}
\end{subequations}
which were derived with the help of
the spin-3/2 generators in their spinor representation,
\begin{subequations}\label{eq:S}
\begin{align}
J_1 &= \frac{\sqrt{3}}{2}
\begin{pmatrix}
  0 & 1 & 0 & 0  \\   1 & 0 & \frac{2}{\sqrt{3}} & 0  \\
0 & \frac{2}{\sqrt{3}} & 0 & 1 \\ 0 & 0 & 1 & 0
  \end{pmatrix} ~,
\\[2ex]
J_2 &= i \frac{\sqrt{3}}{2}
\begin{pmatrix}
  0 & -1 & 0 & 0  \\   1 & 0 & \frac{-2}{\sqrt{3}} & 0  \\
0 & \frac{2}{\sqrt{3}} & 0 & -1 \\ 0 & 0 & 1 & 0
  \end{pmatrix} ~,
 \\[2ex]
J_3 &= \frac{1}{2}
\begin{pmatrix}
  3 & 0 & 0 & 0 \\  0 & 1 & 0 & 0  \\ 0 & 0 & -1 & 0  \\ 0 & 0 & 0 & -3
  \end{pmatrix}
~.
\end{align}
\end{subequations}
For arbitrary spin  of $\Xi$, $K_{ba}$ of Eq.~(\ref{rho32_obs}) becomes
\begin{equation}
  K_{ba} = \frac{\Sigma^{\text{even}}_{ba} - \Sigma^{\text{odd}}_{ba}}
{\Sigma^{\text{even}}_{ba} + \Sigma^{\text{odd}}_{ba}} ~,
\end{equation}
where
\begin{equation}
  \Sigma^{\text{even/odd}}_{ba} = \sum_{m_a,m_b}\frac{d\sigma_{m_b,m_a}}{d\Omega}~,
  \quad \text{$m_a-m_b=\text{even/odd}$~,}
\end{equation}
denotes the sum of all polarized differential cross sections such that the
differences of all possible combinations of initial and final spin projections
$m_a$ and $m_b$ along $\nunit[a]$ and $\nunit[b]$, respectively, are an even or
odd number.

\section{Explicit form of the SDM's} \label{app:SDM_form}

In this section, we list the SDM elements of each $\rho^i$ $(i=0, \ldots, 3)$
in terms of the helicity amplitudes, $\mathcal{H}_i$. The Hermitian $\rho^i$
matrices are arranged according to
\begin{equation}
\hat\rho^i=\begin{pmatrix}
\rho^i_{J,J} & \rho^i_{J,J-1} & \hdots & \rho^i_{J,-J} \\[1ex]
\rho^i_{J-1,J} & \rho^i_{J-1,J-1} & \hdots & \rho^i_{J-1,-J} \\
\vdots & \vdots & \ddots & \vdots & \\
\rho^i_{-J,J} & \rho^i_{-J,J-1} & \hdots &  \rho^i_{-J,-J} \\
\end{pmatrix}  ~.
\end{equation}
For $J=\frac12$, the matrices read explicitly
\begin{subequations}
\begin{align}
{\hat\rho^0} &= \frac 12
\begin{pmatrix}
 \left|\mathcal{H}_1\right|^2+\left|\mathcal{H}_2\right|^2  &
 2i \pi_\Xi^{} \, \mbox{Im} \left[\mathcal{H}_2\mathcal{H}_1^*\right]  \\[1ex]
 -2i \pi_\Xi^{} \, \mbox{Im} \left[\mathcal{H}_2\mathcal{H}_1^*\right]  ~~ &
 \left|\mathcal{H}_1\right|^2 + \left|\mathcal{H}_2\right|^2 \end{pmatrix} ~,
 \\[2ex]
{\hat\rho^1}&=\frac 12
\begin{pmatrix}
2\, \mbox{Re} \left[ \mathcal{H}_2\mathcal{H}_1^* \right] \ \ \ &
\pi_\Xi^{} \left( \left|\mathcal{H}_1\right|^2
- \left|\mathcal{H}_2\right|^2 \right)  \\[1ex]
\pi_\Xi^{} \left( \left|\mathcal{H}_1\right|^2
- \left|\mathcal{H}_2\right|^2 \right)    \ \ \ &
-2\,\mbox{Re}\left[\mathcal{H}_2\mathcal{H}_1^*\right] \end{pmatrix}  ~,
 \\[2ex]
{\hat\rho^2}&=\frac 12
\begin{pmatrix}
 -2\, \mbox{Im}\left[\mathcal{H}_2\mathcal{H}_1^*\right] \ \ \ &
-i \pi_\Xi^{} (\left|\mathcal{H}_1\right|^2+\left|\mathcal{H}_2\right|^2) \\[1ex]
i \pi_\Xi^{} (\left|\mathcal{H}_1\right|^2+\left|\mathcal{H}_2\right|^2)   \ \ \ &
-2\, \mbox{Im}\left[\mathcal{H}_2\mathcal{H}_1^*\right] \end{pmatrix} ~,
 \\[2ex]
{\hat\rho^3}&=\frac 12
\begin{pmatrix}
 \left|\mathcal{H}_1\right|^2-\left|\mathcal{H}_2\right|^2 \ \ \ &
-2\pi_\Xi^{} \, \mbox{Re}\left[\mathcal{H}_2\mathcal{H}_1^*\right]  \\[1ex]
-2\pi_\Xi^{} \, \mbox{Re}\left[\mathcal{H}_2\mathcal{H}_1^*\right]    \ \ \ &
-\left|\mathcal{H}_1\right|^2+\left|\mathcal{H}_2\right|^2 \end{pmatrix}   ~,
\end{align}
\end{subequations}
where $\pi_\Xi^{}$ is the parity of the $\Xi$.

\begin{widetext}
For a $J=\frac32$ resonance,
\begin{subequations}
\begin{align}
\hat\rho^0&=\frac 12
\begin{pmatrix}
\abss{\mathcal{H}_2}+\abss{\mathcal{H}_1} \ \ \ &
\mathcal{H}_2{\mathcal{H}_4}^*+\mathcal{H}_1{\mathcal{H}_3}^* \ \ \ &
\pi_\Xi^{} \left(\mathcal{H}_1{\mathcal{H}_4}^*-\mathcal{H}_2{\mathcal{H}_3}^* \right) \ \ \ &
2i\pi_\Xi^{} \, \mbox{Im} \left[\mathcal{H}_2{\mathcal{H}_1}^* \right] \\[1ex]
\mathcal{H}_4{\mathcal{H}_2}^*+\mathcal{H}_3{\mathcal{H}_1}^* \ \ \ & \abss{\mathcal{H}_4}+\abss{\mathcal{H}_3} \ \ \ &
2i \pi_\Xi^{}\, \mbox{Im}\left[\mathcal{H}_3{\mathcal{H}_4}^*\right] \ \ \ &
\pi_\Xi^{} \left(-\mathcal{H}_3{\mathcal{H}_2}^*+\mathcal{H}_4{\mathcal{H}_1}^*\right) \\[1ex]
\pi_\Xi^{} \left(\mathcal{H}_4{\mathcal{H}_1}^*-\mathcal{H}_3{\mathcal{H}_2}^* \right) \ \ \ &
-2i \pi_\Xi^{}\, \mbox{Im}\left[\mathcal{H}_3{\mathcal{H}_4}^*\right]  \ \ \ & \abss{\mathcal{H}_4}+\abss{\mathcal{H}_3} \ \ \ &
-\mathcal{H}_4{\mathcal{H}_2}^*-\mathcal{H}_3{\mathcal{H}_1}^* \\[1ex]
-2i\pi_\Xi^{} \, \mbox{Im} \left[\mathcal{H}_2{\mathcal{H}_1}^* \right] \ \ \ &\pi_\Xi^{} \left(-\mathcal{H}_2{\mathcal{H}_3}^*+\mathcal{H}_1{\mathcal{H}_4}^*\right) \ \ \ &
-\mathcal{H}_2{\mathcal{H}_4}^*-\mathcal{H}_1{\mathcal{H}_3}^* \ \ \ & \abss{\mathcal{H}_2}+\abss{\mathcal{H}_1} \end{pmatrix}  ~,
 \\[2ex]
 \hat\rho^1&=\frac 12
\begin{pmatrix}
2\, \mbox{Re}\left[\mathcal{H}_2{\mathcal{H}_1}^*\right] \ \ \ &
\mathcal{H}_1{\mathcal{H}_4}^*+\mathcal{H}_2{\mathcal{H}_3}^* \ \ \ &
\pi_\Xi^{} \left(\mathcal{H}_2{\mathcal{H}_4}^*-\mathcal{H}_1{\mathcal{H}_3}^*\right) \ \ \ &
 \pi_\Xi\left(-\abss{\mathcal{H}_2}+\abss{\mathcal{H}_1}\right) \\[1ex]
\mathcal{H}_4{\mathcal{H}_1}^*+\mathcal{H}_3{\mathcal{H}_2}^* \ \ \ & 2\, \mbox{Re}\left[\mathcal{H}_4{\mathcal{H}_3}^*\right] \ \ \ &
\pi_\Xi^{} \left(\abss{\mathcal{H}_4}-\abss{\mathcal{H}_3}\right) \ \ \ &
\pi_\Xi^{} \left(-\mathcal{H}_4{\mathcal{H}_2}^*+\mathcal{H}_3{\mathcal{H}_1}^* \right)\\[1ex]
\pi_\Xi^{} \left(\mathcal{H}_4{\mathcal{H}_2}^*-\mathcal{H}_3{\mathcal{H}_1}^*\right) \ \ \ &
\pi_\Xi^{} \left(\abss{\mathcal{H}_4}-\abss{\mathcal{H}_3}\right) \ \ \ & -2\, \mbox{Re}\left[\mathcal{H}_4{\mathcal{H}_3}^*\right] \ \ \ &
\mathcal{H}_3{\mathcal{H}_2}^*+\mathcal{H}_4{\mathcal{H}_1}^* \\[1ex]
 \pi_\Xi\left(-\abss{\mathcal{H}_2}+\abss{\mathcal{H}_1}\right) \ \ \ &\pi_\Xi^{} \left(-\mathcal{H}_2{\mathcal{H}_4}^*+\mathcal{H}_1{\mathcal{H}_3}^* \right) \ \ \ &
 \mathcal{H}_2{\mathcal{H}_3}^*+\mathcal{H}_1{\mathcal{H}_4}^* \ \ \ & -2\, \mbox{Re}\left[\mathcal{H}_2{\mathcal{H}_1}^*\right] \end{pmatrix}  ~,
 \\[2ex]
  \hat\rho^2&=\frac 12
\begin{pmatrix}
2 \, \mbox{Im}\left[\mathcal{H}_1{\mathcal{H}_2}^*\right] \ \ \ &
i\left(-\mathcal{H}_1{\mathcal{H}_4}^*+\mathcal{H}_2{\mathcal{H}_3}^*\right) \ \ \ &
i\pi_\Xi^{} \left(\mathcal{H}_2{\mathcal{H}_4}^*+\mathcal{H}_1{\mathcal{H}_3}^*\right) \ \ \ &
-i\pi_\Xi^{} \left(\abss{\mathcal{H}_2}+\abss{\mathcal{H}_1}\right) \\[1ex]
i\left(\mathcal{H}_4{\mathcal{H}_1}^*-\mathcal{H}_3{\mathcal{H}_2}^*\right) \ \ \ & 2 \, \mbox{Im}\left[\mathcal{H}_3{\mathcal{H}_4}^*\right] \ \ \ &
i\pi_\Xi^{} \left(\abss{\mathcal{H}_4}+\abss{\mathcal{H}_3}\right) \ \ \ &
-i\pi_\Xi^{} \left(\mathcal{H}_4{\mathcal{H}_2}^*+\mathcal{H}_3{\mathcal{H}_1}^*\right) \\[1ex]
-i\pi_\Xi^{} \left(\mathcal{H}_4{\mathcal{H}_2}^*+\mathcal{H}_3{\mathcal{H}_1}^*\right)  \ \ \ &-i\pi_\Xi^{} \left(\abss{\mathcal{H}_4}+\abss{\mathcal{H}_3}\right) \ \ \ &
 2\, \mbox{Im}\left[\mathcal{H}_3{\mathcal{H}_4}^*\right] \ \ \ &
i\left(\mathcal{H}_3{\mathcal{H}_2}^*-\mathcal{H}_4{\mathcal{H}_1}^*\right) \\[1ex]
i\pi_\Xi^{} \left(\abss{\mathcal{H}_2}+\abss{\mathcal{H}_1}\right) \ \ \ & i\pi_\Xi^{} \left(\mathcal{H}_2{\mathcal{H}_4}^*+\mathcal{H}_1{\mathcal{H}_3}^*\right) \ \ \ &
i\left(-\mathcal{H}_2{\mathcal{H}_3}^*+\mathcal{H}_1{\mathcal{H}_4}^*\right) \ \ \ & 2\, \mbox{Im}\left[\mathcal{H}_1{\mathcal{H}_2}^*\right] \end{pmatrix} ~,
 \\[2ex]
\hat\rho^3&=\frac 12
\begin{pmatrix}
-\abss{\mathcal{H}_2}+\abss{\mathcal{H}_1}  \ \ \ &
-\mathcal{H}_2{\mathcal{H}_4}^*+\mathcal{H}_1{\mathcal{H}_3}^* \ \ \ &
\pi_\Xi^{} \left(\mathcal{H}_1{\mathcal{H}_4}^*+\mathcal{H}_2{\mathcal{H}_3}^*\right) \ \ \ &
-2\pi_\Xi^{} \, \mbox{Re}\left[\mathcal{H}_2{\mathcal{H}_1}^*\right] \\[1ex]
-\mathcal{H}_4{\mathcal{H}_2}^*+\mathcal{H}_3{\mathcal{H}_1}^* \ \ \ & -\abss{\mathcal{H}_4}+\abss{\mathcal{H}_3} \ \ \ &
2\pi_\Xi^{} \, \mbox{Re}\left[\mathcal{H}_4{\mathcal{H}_3}^*\right] \ \ \ &
\pi_\Xi^{} \left(-\mathcal{H}_3{\mathcal{H}_2}^*-\mathcal{H}_4{\mathcal{H}_1}^* \right) \\[1ex]
\pi_\Xi^{} \left(\mathcal{H}_4{\mathcal{H}_1}^*+\mathcal{H}_3{\mathcal{H}_2}^*\right) \ \ \ &2\pi_\Xi^{} \, \mbox{Re}\left[\mathcal{H}_4{\mathcal{H}_3}^*\right] \ \ \ &
 \abss{\mathcal{H}_4}-\abss{\mathcal{H}_3} \ \ \ & -\mathcal{H}_4{\mathcal{H}_2}^*+\mathcal{H}_3{\mathcal{H}_1}^* \\[1ex]
-2\pi_\Xi^{} \, \mbox{Re}\left[\mathcal{H}_2{\mathcal{H}_1}^*\right] \ \ \ & \pi_\Xi^{} \left(-\mathcal{H}_2{\mathcal{H}_3}^*-\mathcal{H}_1{\mathcal{H}_4}^* \right) \ \ \ &
  -\mathcal{H}_2{\mathcal{H}_4}^*+\mathcal{H}_1{\mathcal{H}_3}^* \ \ \ & \abss{\mathcal{H}_2}-\abss{\mathcal{H}_1} \end{pmatrix} \ ~.
 \end{align}
\end{subequations}
\end{widetext}

\section{Measuring the SDM Elements} \label{app:measure_SDM}

In Sec.~\ref{sec:SDMA}, we identified a set of SDM elements that determines the
reaction amplitude completely. A standard way of measuring the SDM elements is
via the subsequent decay of the produced particle in a primary reaction by
exploiting the self-analyzing property of the decay-product particle.

In the present work, the reaction in Eq.~(\ref{eq:Kbar-1}) is the primary (or
production) reaction, where the $\Xi$ baryon is produced. If the produced $\Xi$
is a ground state $\Xi$, then, it decays via a single weak decay process into
\begin{equation}
\Xi \to \Lambda + \pi  ,
\label{eq:Xi-sd}
\end{equation}
whose associated $\Xi$ decay-asymmetry parameters, $\alpha_\Xi^{}$, are known
to be~\cite{PDG12}
\begin{subequations}\label{eq:decay-asymmXi}
\begin{align}
\alpha_{\Xi^0}^{}  &= -0.406 \pm 0.013~,
\\
\alpha_{\Xi^-}^{}  &= -0.458 \pm 0.012  ~.
\end{align}
\end{subequations}

An excited $\Xi$ resonance, $\Xi^*$, on the other hand, may undergo a
double-decay process
\begin{subequations}\label{eq:Xi-dd}
\begin{align}
\Xi^* &\rightarrow\Xi+\pi \nonumber \\
&\hspace{1.7em} \DRarrow \Lambda+\pi  ~,
 \\
 \intertext{or}
 \Xi^*&\rightarrow\Lambda+\bar{K} \nonumber \\
&\hspace{1.7em} \DRarrow N+\pi  ~.
\end{align}
\end{subequations}
The associated $\Lambda$ decay-asymmetry parameter for the second-step process
$\Lambda \to N+\pi$ is~\cite{PDG12}
\begin{align}
\alpha_{\Lambda -}^{}  & = +0.642 \pm 0.013  \qquad  (\Lambda^0 \to p + \pi^-)  ~,
\nonumber \\
\alpha_{\Lambda 0}^{} & = +0.650 \pm 0.015 \qquad (\Lambda^0 \to n + \pi^0)  ~.
\label{eq:decay-asymmLa}
\end{align}

The $\Xi$ production process (\ref{eq:Kbar-1}) is described in the
CM frame of the reaction. The $\Xi$ decay processes of
Eqs.~(\ref{eq:Xi-sd}) and (\ref{eq:Xi-dd}), on the other hand, are described in
the rest frame of the produced $\Xi$, whose right-handed
Cartesian coordinate system $\{\nunit'_1, \nunit'_2\equiv\nunit^{}_2,
\nunit'_3\equiv\nunit[p]' \}$ is fully specified in Fig.~\ref{fig:1}.

In the double-decay processes (\ref{eq:Xi-dd}), the subsequent $\Lambda$ decay
process is described in the rest frame of the decaying $\Lambda$ denoted by
$\{\nunit''_1, \nunit''_2,\nunit''_3\}$, with $\nunit''_3\equiv
\nunit[p]^{}_\Lambda$, where $\bm{p}^{}_\Lambda$ describes the direction of the
$\Lambda$'s momentum in the $\{\nunit'_1, \nunit'_2,\nunit'_3\}$ frame (see
Fig.~\ref{fig:1}); the other two axes are given by
$\nunit''_2=(\nunit'_3\times\nunit[p]^{}_\Lambda)/|\nunit'_3\times\nunit[p]^{}_\Lambda|$
and $\nunit''_1=\nunit''_2\times\nunit''_3$.

\subsection{\boldmath Single-decay process: Ground state $\Xi$}

 The ground-state $\Xi$ decays weakly almost entirely into $\Lambda+\pi$.
We define the amplitude describing the $\Xi$ production process
$\bar{K} + N \to K + \Xi$, followed by the subsequent weak decay of the
produced $\Xi$, $\Xi \to \Lambda + \pi$, as~\cite{BBBB87b, Chung71}

\begin{align}
A &\equiv A(\Omega_\Xi,\Omega_\Lambda,\lambda_\Xi,\lambda_\Lambda,\lambda_N)
\nonumber \\
& = \bra{\Omega_\Lambda,\lambda_\Lambda} \hat M_D \ket{\lambda_\Xi}
\bra{\Omega_\Xi,\lambda_\Xi} \hat M \ket{\lambda_N}
\label{eq:A_PD}
\end{align}
with $\bra{\Omega_\Xi,\lambda_\Xi} \hat M \ket{\lambda_N}$ denoting the
production reaction amplitude (in the corresponding CM frame) and
\begin{equation}
\bra{\Omega_\Lambda,\lambda_\Lambda} \hat M_D \ket{\lambda_\Xi}
\equiv \sqrt{\frac{2 J+1}{4\pi}} F^{\Xi}_{\lambda_\Lambda}
D^{J*}_{\lambda_\Xi,\lambda_\Lambda}(\Omega_\Lambda) ,
\end{equation}
denoting the subsequent $\Xi$ decay amplitude (in the rest frame of $\Xi$).
Here, $\Omega_\Xi$ and $\Omega_\Lambda$ are short-hand notations, respectively,
for the polar and azimuthal angles of the produced $\Xi$ in the CM frame of the
production, $\Omega_\Xi=(\theta,\phi=0)$, and for the polar and azimuthal
angles of the $\Lambda$ in the rest frame of the produced $\Xi$,
$\Omega_\Lambda=(\theta_\Lambda,\phi_\Lambda)$.
$\lambda_\Xi$($\lambda_\Lambda$) is the helicity of the $\Xi$($\Lambda$) in the
respective frame, while $J$ denotes the spin of the decaying $\Xi$.
$F^\Xi_{\lambda_\Lambda}$ stands for the helicity $\Xi$-decay amplitude and
$D^J_{\lambda_\Xi,\lambda_\Lambda}(\Omega_\Lambda)$ is the usual Wigner
rotation matrix. Here, the argument $\Omega_\Lambda$ in
$D^J_{\lambda_\Xi,\lambda_\Lambda}(\Omega_\Lambda)$ is to be understood as the
set of Euler angles $\{\alpha, \beta, \gamma\}$, such that,
$D^J_{\lambda_\Xi,\lambda_\Lambda}(\Omega_\Lambda) \equiv
D^J_{\lambda_\Xi,\lambda_\Lambda}(\alpha=\phi_\Lambda, \beta=\theta_\Lambda,
\gamma =0)$ in the conventions defined in Ref.~\cite{Edmonds}.

The angular distribution of the $\Lambda$ hyperon, $I(\Omega_\Lambda)$, in the
$\Xi \to \Lambda + \pi$ decay (for fixed $\Xi$ angle $\Omega_\Xi$) is given by
\begin{equation}
I(\Omega_\Lambda) = \sum_{i=0}^3 P_{i}\, I^i(\Omega_\Lambda) \ ,
\label{eq:decay-ang0}
\end{equation}
where
\begin{widetext}
\begin{align}
I^i(\Omega_\Lambda) & \equiv \sum_{\rm{all\ \lambda's}}
A(\Omega_\Xi,\Omega_\Lambda,\lambda_\Xi,\lambda_\Lambda,\lambda_N) \,
\rho^{N,i}_{\lambda_N,\lambda'_N}
A^*(\Omega_\Xi,\Omega_\Lambda,\lambda'_\Xi,\lambda_\Lambda,\lambda'_N)
\nonumber \\
&=\frac{(2 J+1)}{4\pi} \sum_{\rm all\ \lambda's}
F^{\Xi}_{\lambda_\Lambda} F^{\Xi*}_{\lambda_\Lambda}\,
M_{\lambda_\Xi,\lambda_N} \, \rho^{N,i}_{\lambda_N,\lambda'_N} \,
M^*_{\lambda'_\Xi,\lambda'_N}
D^{J*}_{\lambda_\Xi,\lambda_\Lambda}(\Omega_\Lambda)
D^{J}_{\lambda'_\Xi,\lambda'_\Lambda}(\Omega_\Lambda)
\nonumber \\
&=\frac{(2J+1)}{4\pi} \sum_{\rm all\ \lambda's}
F^{\Xi}_{\lambda_\Lambda} F^{\Xi*}_{\lambda_\Lambda}
\, \rho^{\Xi,i}_{\lambda_\Xi,\lambda'_\Xi} \,
D^{J*}_{\lambda_\Xi,\lambda_\Lambda}(\Omega_\Lambda)
D^{J}_{\lambda'_\Xi,\lambda'_\Lambda}(\Omega_\Lambda)  ,
\label{eq:decay-ang1}
\end{align}
\end{widetext}
Here,
$\rho^{N,i}_{\lambda_N,\lambda_N'}=\braket{\lambda_N|\hat\rho^{N,i}|\lambda_N'}$
denotes the target nucleon SDM element with $\hat\rho^{N,i}$ given by
Eq.~(\ref{rhoN_def1}), and Eq.~(\ref{eq:SDM-XiN}) was used in the last step.
Also, we note that the explicit reference to the $\Omega_\Xi$ dependence of the
angular distribution $I^i$ in Eqs.~(\ref{eq:decay-ang0}) and
(\ref{eq:decay-ang1}) has been suppressed for the sake of simplicity of
notation. The same holds for the angular distribution in
Eqs.~(\ref{eq:2decay-ang0}) and (\ref{eq:D2}) in the next subsection.

We now define the moments, $H^i(L,M)$, of this distribution as
\begin{align}
H^i(L,M) &\equiv \int d\Omega_\Lambda\, I^i(\Omega_\Lambda)
D^L_{M,0}(\Omega_\Lambda)
\nonumber \\
&= t^{J,i}_{LM} \sum_{\lambda_\Lambda} F^{\Xi}_{\lambda_\Lambda}
F^{\Xi*}_{\lambda_\Lambda}
\braket{ J \, \lambda_\Lambda \, L \, 0 | J \, \lambda_\Lambda }  ,
\label{eq:moments}
\end{align}
where $d\Omega_\Lambda \equiv \sin\theta_\Lambda\, d\theta_\Lambda\,
d\phi_\Lambda$. The quantity
$t^{J,i}_{LM}$ here is related to the SDM elements of the $\Xi$
by~\cite{Chung71}
\begin{eqnarray}
t^{J,i}_{LM} &\equiv& \sum_{\lambda_\Xi,\lambda'_\Xi}
\rho^{\Xi,i}_{\lambda_\Xi,\lambda'_\Xi}
\braket{ J \, \lambda'_\Xi \, L \, M | J \, \lambda_\Xi }  ~,
\label{tJLM_def}
\end{eqnarray}
whose inversion produces
\begin{equation}
\rho^{\Xi,i}_{\lambda_\Xi,\lambda_\Xi'} = \sum_{L}
\frac{2L+1}{2J+1}
\braket{ J \, \lambda_\Xi' \, L \, M | J \, \lambda_\Xi }
t^{J,i}_{LM}  ~,
\label{rho_t}
\end{equation}
where $M=\lambda^{}_\Xi-\lambda'_\Xi$.

Introducing further the quantities
\begin{equation}
g^\Xi_\pm  \equiv F^{\Xi}_{\pm\frac12} F^{\Xi*}_{\pm\frac12}   ~,
\label{eq:gXi}
\end{equation}
we can re-express the moments in Eq.~(\ref{eq:moments}) as
\begin{equation}
H^i(L,M) =t^{J,i}_{LM}\; \braket{ J \, \tfrac12 \, L \, 0 | J\, \tfrac12 }
\left[ g^\Xi_+ + (-1)^L g^\Xi_-\right] \ .
\end{equation}
We note that $g^\Xi_\pm$ introduced in Eq.~(\ref{eq:gXi}) are related to the
$\Xi$ decay-asymmetry parameter, $\alpha_\Xi^{}$, given in
Eq.~(\ref{eq:decay-asymmXi}) by~\cite{Kallen,Gasiorowicz}
\begin{equation}
\frac{g^\Xi_+-g^\Xi_-}{g^\Xi_++g^\Xi_-}=\alpha_\Xi^{} ~.
\end{equation}
Then, taking the ratio of the moments $H^i(L,M)$ $(i=1,2,3)$ and $H^0(0,0)$,
we obtain
\begin{equation}
\frac{H^i(L,M)}{H^0(0,0)}  = \zeta_L^{} \, \frac{t^{J,i}_{LM}}{t^{J,0}_{00}}
\braket{ J \, \tfrac12 \, L \, 0 | J \, \tfrac12 } \ ,
\label{HLM_1hf}
\end{equation}
where $\zeta_L^{} = 1$ for even $L$ and $\zeta_L^{} = \alpha_\Xi^{}$ for odd $L$.

Now, from Eq.~(\ref{tJLM_def}) and the definition of $\hat{\rho}^{\Xi, 0}$ in
Eq.~(\ref{rhoXii_def}), we get
\begin{equation}
t^{J,0}_{00} = \mbox{Tr} \left[ \hat{\rho}^{\Xi, 0} \right]
= \frac{d\sigma}{d\Omega}  ~.
\label{eq:t00-norm}
\end{equation}
One can now use Eq.~(\ref{HLM_1hf}) to
extract $t^{J,i}_{LM}$. Once $t^{J,i}_{LM}$ is known, the SDM elements
$\rho^{\Xi,i}_{\lambda_\Xi,\lambda_\Xi'}$ are obtained by making use of
Eq.~(\ref{rho_t}). Note that the non-vanishing moments $H^i(LM)$ are restricted
to $L \le 2J$ and $|M| \le L$.

\subsection{\boldmath Double-decay process: Excited $\Xi$ resonance}

The double-decay processes shown in Eq.~(\ref{eq:Xi-dd}) are treated
analogously to the single-decay process of the previous subsection. Here, we
discuss the decay chain with the subsequent decay of the $\Lambda$ hyperon,
$\Lambda \to p + \pi^-$, but the results apply to any decay chain that is a
strong decay followed by a weak decay and containing a single pseudoscalar
meson at each step of the decay.

As for the single-decay process case discussed in the previous subsection, we
begin by defining the amplitude describing the $\Xi^*$ production process
$\bar{K} + N \to K + \Xi^*$, followed by the strong decay of the produced $\Xi^*$,
$\Xi^* \to \Lambda + K$, and the subsequent weak decay of $\Lambda$,
$\Lambda \to N + \pi$, as~\cite{Chung71,BBBB87b}

\begin{align}
A &\equiv A(\Omega_\Xi,\Omega_\Lambda,\Omega_p,
\lambda_N,\lambda_\Xi,\lambda_\Lambda,\lambda_p)
\nonumber \\
&=\braket{\Omega_p,\lambda_p | \hat M'_D | \lambda_\Lambda}
\braket{\Omega_\Lambda,\lambda_\Lambda | \hat M_D | \lambda_\Xi}
\braket{\Omega_\Xi,\lambda_\Xi | \hat M | \lambda_N} ~,
\label{A_PDD}
\end{align}
where $\bra{\Omega_\Xi,\lambda_\Xi} \hat M \ket{\lambda_N}$ stands for the
$\Xi^*$ production amplitude and
\begin{subequations}
\begin{align}
\bra{\Omega_\Lambda,\lambda_\Lambda} \hat M_D \ket{\lambda_\Xi} &\equiv
\sqrt{\frac{2J+1}{4\pi}} \, F^{\Xi}_{\lambda_\Lambda}
D^{J*}_{\lambda_\Xi,\lambda_\Lambda}(\Omega_\Lambda) ~,
\ \\
\bra{\Omega_p,\lambda_p} \hat M'_D \ket{\lambda_\Lambda} &\equiv
\sqrt{\frac{2}{4\pi}} \, F^{\Lambda}_{\lambda_p}
D^{\frac12*}_{\lambda_\Lambda,\lambda_p}(\Omega_p) ~ ,
\end{align}
\end{subequations}
denote the subsequent $\Xi^*$ strong-decay and $\Lambda$ weak-decay amplitudes,
respectively. We note that the $\Xi$ production and decay amplitudes are
calculated in the CM frame of the production reaction and the rest frame of the
produced $\Xi$, respectively, exactly in the same way as for the single-decay
case discussed in the previous subsection. The subsequent $\Lambda$-decay
amplitude, $\bra{\Omega_p,\lambda_p} \hat M'_D \ket{\lambda_\Lambda}$, is
calculated in the rest frame of the decaying $\Lambda$ denoted by $\{
\nunit''_1,\nunit''_2,\nunit''_3 \}$ [cf.\ the second paragraph just below
Eq.~(\ref{eq:decay-asymmLa})], where $\Omega_p=(\theta_p,\phi_p)$ is a
short-hand notation for the polar and azimuthal angles $\theta_p$ and $\phi_p$,
respectively, of the decay-product proton measured in the $\Lambda$ rest frame.

The angular distribution of the entire double decay process (for fixed $\Xi$
production angle $\Omega_\Xi$) is given as
\begin{equation}
I(\Omega_\Lambda,\Omega_p) =
\sum_{i=0}^3 P_{i} I^i(\Omega_\Lambda,\Omega_p) ~ ,
\label{eq:2decay-ang0}
\end{equation}
where
\begin{widetext}
\begin{align}
I^i\left(\Omega_\Lambda,\Omega_p\right) & \equiv
\sum_{\mathrm{all\ \lambda's}}
A(\Omega_\Xi,\Omega_\Lambda,\Omega_p,
\lambda_N,\Lambda_\Xi,\lambda_\Lambda,\lambda_p) \,
\rho^{N,i}_{\lambda_N,\lambda_N'}
A^*(\Omega_\Xi,\Omega_\Lambda,\Omega_p,
\lambda'_N,\lambda'_\Xi,\lambda'_\Lambda,\lambda_p)
\nonumber \\
&=\frac{2(2J+1)}{16\pi^2}
\sum_{\mathrm{all\ \lambda's}}
\rho^{\Xi,i}_{\lambda_\Xi,\lambda'_\Xi}\,
g^\Lambda_{\lambda_p}\,g^\Xi_{\lambda_\Lambda,\lambda'_\Lambda}
D^{\frac 12*}_{\lambda_\Lambda,\lambda_p}(\Omega_p)
D^{\frac 12}_{\lambda'_\Lambda,\lambda_p}(\Omega_p)
D^{\mathrm{J}*}_{\lambda_\Xi,\lambda_\Lambda}(\Omega_\Lambda)
D^{\mathrm{J}}_{\lambda'_\Xi,\lambda'_\Lambda}(\Omega_\Lambda)  ~,
\label{eq:D2}
\end{align}
with
\begin{equation}
g^\Lambda_{\lambda_p} \equiv
F^\Lambda_{\lambda_p}\, F^{\Lambda*}_{\lambda_p}  \quad\mbox{ and }\quad
g^\Xi_{\lambda_\Lambda,\lambda'_\Lambda} \equiv
F^\Xi_{\lambda_\Lambda}\, F^{\Xi*}_{\lambda'_\Lambda}  ~.
\end{equation}
To arrive at the last equality in Eq.~(\ref{eq:D2}), we have made use of
Eq.~(\ref{eq:SDM-XiN}).

We now define the moments $H^i(l,m,L,M)$ as
\begin{align}
H^i(l,m,L,M) & \equiv \int d\Omega_\Lambda d\Omega_p
I^i(\Omega_\Lambda,\Omega_p) \, D^L_{M,m}(\Omega_\Lambda) D^l_{m,0}(\Omega_p)
\nonumber \\
&= t^{\mathrm{J},i}_{LM}
\sum_{\lambda_\Lambda,\lambda'_\Lambda} g^\Xi_{\lambda_\Lambda,\lambda'_\Lambda}
\braket{ J \, \lambda'_\Lambda \, L \, m | J \, \lambda_\Lambda }
\braket{ \tfrac12 \, \lambda'_\Lambda \, l \, m | \tfrac12 \, \lambda_\Lambda }
 \sum_{\lambda_p}
g^\Lambda_{\lambda_p}
\braket{ \tfrac12 \, \lambda_p \, l \, 0 | \tfrac12 \, \lambda_p } ~,
\label{HlmLM}
\end{align}
with $t^{J,i}_{LM}$ given by Eq.~(\ref{tJLM_def}).
\end{widetext}

The different $g^\Xi_{\lambda_\Lambda,\lambda_\Lambda'}$ are related to each
other by
\begin{subequations}\label{g_Xi}
\begin{align}
g^\Xi_{--} &= g^\Xi_{++} ~,
 \\
g^\Xi_{+-} &= g^\Xi_{-+} = \pi_\Xi^{} \, (-1)^{J+\frac12} \, g^\Xi_{++}  ~.
\end{align}
\end{subequations}
The $g^\Lambda_{\lambda_p}$ terms can be related to the $\Lambda$ decay
asymmetry parameter, $\alpha_\Lambda^{}$, by~\cite{Kallen,Gasiorowicz}
\begin{equation}
\frac{g^\Lambda_+ - g^\Lambda_-}{g^\Lambda_+ + g^\Lambda_-}
= \alpha_\Lambda^{} ~.
\end{equation}

Note that the non-vanishing moments $H^i(l,m,L,M)$ are restricted to $|m| \le
l$, $l \le 1$, $|M| \le L$ and $L \le 2J$, as can be read off from
Eq.~(\ref{HlmLM}). The moments $H^i(0,0,L,M)$ and $H^i(1,m,L,M)$ vanish
identically for odd and even $L$, respectively, due to Eqs.~(\ref{HlmLM}) and
(\ref{g_Xi}). Analogously to the single-decay case, the ratios of the moments
\begin{subequations}\label{HLM_3hf}
\begin{align}
\frac{H^i(0,0,L,M)}{H^0(0,0,0,0)} &=
\frac{t^{J,i}_{LM}}{t^{J,0}_{00}} \,
\braket{ J \, \tfrac12 \, L \, 0 | J \, \tfrac12 }
\\
\intertext{for even $L$ and}
 \frac{H^i(1,0,L,M)}{H^0(0,0,0,0)} &= \frac{\alpha^{}_\Lambda}{3}\,
\frac{t^{J,i}_{LM}}{t^{J,0}_{00}} \,
\braket{ J \, \tfrac12 \, L \, 0 | J \, \tfrac12 }
\end{align}
\end{subequations}
for odd $L$  allow us to determine $t^{J,i}_{LM}$. Since $t^{J,0}_{00} =
{d\sigma}/{d\Omega}$, once $t^{J,i}_{LM}$ is extracted, the SDM elements
$\rho^{\Xi,i}_{\lambda_\Xi,\lambda_\Xi'}$ can be determined via
Eq.~(\ref{rho_t}).

\end{document}